\documentclass[11pt]{article}

\usepackage[letterpaper,margin=1in]{geometry}
\usepackage[T1]{fontenc}
\usepackage[utf8]{inputenc}
\usepackage{amsmath,amssymb,amsthm,bm}
\usepackage{graphicx}
\usepackage{xcolor}
\usepackage{xparse}
\usepackage{tikz}
\usepackage{url}
\usepackage[numbers,sort&compress]{natbib}

\makeatletter

\renewcommand{\bibsection}{\section*{References}}
\makeatother
\usepackage[acronym]{glossaries}
\glsdisablehyper
\usepackage[colorlinks=true,linkcolor=blue,citecolor=blue,urlcolor=blue]{hyperref}
\usepackage{cleveref}

\definecolor{lavender}{RGB}{212,197,199}

\newcommand{\fsq}[1]{%
  \tikz[baseline=-0.0001ex]\draw[fill=#1,draw=black] (0,0) rectangle (1ex,1ex);}

\newcommand{\fcirc}[1]{%
  \tikz[baseline=-0.0001ex]\draw[fill=#1,draw=black] (0.6ex,0.6ex) circle (0.6ex);}

\NewDocumentCommand{\circmark}{O{white}}{%
  \tikz[baseline=-0.5ex]%
    \node[draw=lavender, fill=#1, circle,
          inner sep=0pt, minimum size=1.2ex, line width=0.4pt] (c) {};%
}

\NewDocumentCommand{\sqmark}{O{white}}{%
  \tikz[baseline=-0.5ex]%
    \node[draw=lavender, fill=#1,
          inner sep=0pt, minimum size=1.2ex, line width=0.4pt] (s) {};%
}

\newacronym{RMFR}{RMFR}{reverse-magnetic-field reciprocity}
\newacronym{DC}{DC}{direct current}
\newacronym{AC}{AC}{alternating current}
\newacronym{VdP}{VdP}{Van der Pauw}
\newacronym{ZFC}{ZFC}{zero-field cooled}
\newacronym{EB}{EB}{exchange-biased}
\newacronym{AHE}{AHE}{anomalous Hall effect}
\newacronym{THE}{THE}{topological Hall effect}
\newacronym{SSC}{SSC}{scalar spin chirality}
\newacronym{TMDs}{TMDs}{transition metal dichalcogenides}
\newacronym{TRS}{TRS}{time-reversed states}

\title{Measuring the Hall Effect in Hysteretic Materials}

\author{%
Jaime M. Moya$^{1}$, Anthony Voyemant$^{2}$, Sudipta Chatterjee$^{1}$, Scott B. Lee$^{1}$,\\
Grigorii Skorupskii$^{1}$, Connor J. Pollak$^{1}$, and Leslie M. Schoop$^{1}$\\[0.8em]
\small $^{1}$Department of Chemistry, Princeton University, Princeton, New Jersey, USA\\
\small $^{2}$Department of Physics and Astronomy, California State University, Northridge, California, USA\\[0.8em]
\small Correspondence: Leslie M. Schoop (\href{mailto:lschoop@princeton.edu}{lschoop@princeton.edu})
}

\date{}

\begin{document}
\maketitle

\begin{abstract}
Measurement of the Hall effect is a ubiquitous probe for materials discovery, characterization, and metrology. Inherent to the Hall measurement geometry, the measured signal is often contaminated by unwanted contributions, so the data must be processed to isolate the Hall response. The standard approach invokes Onsager--Casimir reciprocity and antisymmetrizes the raw signal about zero applied magnetic field. In hysteretic materials this becomes nontrivial, since Onsager--Casimir relations apply only to microscopically reversible states. Incorrect antisymmetrization can lead to artifacts that mimic anomalous or topological Hall signatures. The situation is especially subtle when hysteresis loops are not centered at zero applied field, as in exchange-biased systems. A practical reference for generically extracting the Hall response in hysteretic materials is lacking. Here, using Co$_3$Sn$_2$S$_2$ as a bulk single-crystal model that can be prepared with or without exchange-biased hysteresis, we review and demonstrate two procedures that can be used to extract the Hall effect: (1) reverse-magnetic-field reciprocity and (2) antisymmetrization with respect to applied field. We then measure the Hall effect on CeCoGe$_3$, a noncentrosymmetric antiferromagnet which can be prepared to have asymmetric magnetization and magnetoresistance, and demonstrate how improper processing can generate artificial anomalous Hall signals. The methods reviewed are generic and can be applied to any conductor.

\end{abstract}


\vspace{0.8em}
\noindent\textbf{Abbreviations:} RMFR, reverse-magnetic-field reciprocity; DC, direct current; AC, alternating current; VdP, Van der Pauw; ZFC, zero-field cooled; EB, exchange-biased; AHE, anomalous Hall effect; THE, topological Hall effect; SSC, scalar spin chirality; TMDs, transition metal dichalcogenides; TRS, time-reversed states.

\section{Introduction}\label{Introduction}

The Hall effect refers to the development of a transverse voltage across a material carrying current in the presence of a magnetic field via the antisymmetric component of the resistivity tensor. It provides crucial insights into charge carrier properties such as type, density, and mobility, making it a foundational tool in materials characterization. Beyond traditional applications, the Hall effect underpins key advances in spintronics \cite{jeon2024multicore,vsmejkal2022anomalous,vsmejkal2018topological}, topological materials \cite{liu2016quantum}, and sensor technologies \cite{karsenty2020comprehensive}. In the materials discovery community, measurements of the Hall effect have been extensively used in recent years to identify and/or characterize systems with reciprocal-space Berry curvature by the \gls{AHE} \cite{chang2013experimental,deng2020quantum, nagaosa2010anomalous,xiao2010berry,liu2016quantum,vsmejkal2022anomalous}, real-space Berry curvature by the \gls{THE} \cite{neubauer2009topological,vsmejkal2018topological}, and complex spin textures with non-zero \gls{SSC} \cite{taguchi2001spin,vsmejkal2018topological}.  With the proliferation of reports of such exotic Hall responses, ensuring accurate and reproducible Hall measurements has become increasingly important. However, in materials that exhibit magnetic hysteresis or complex domain behavior, extracting the true Hall signal can be nontrivial and may lead to artificial features that mimic \gls{AHE}-, \gls{THE}-, or \gls{SSC}-like signatures.

The practical difficulty is that the measured transverse signal often contains unwanted contributions due to contact misalignment or other factors, so the data must be processed to isolate the Hall response. In non-hysteretic materials, this is usually done by antisymmetrizing the signal at equal and opposite applied field. In hysteretic materials, however, this procedure becomes nontrivial because Onsager--Casimir \cite{onsager1931reciprocal,onsager1931reciprocal2,casimir1945onsager} reciprocity applies between \gls{TRS} at opposite magnetic induction fields, which do not necessarily correspond to equal and opposite applied fields during a continuous sweep. If this distinction is ignored, artificial \gls{AHE}- or \gls{THE}-like features can result.

While practical guidance exists for Hall measurements in non-magnetic materials \cite{lindemuth2020hall}, the distinction between applied field and magnetic induction field is usually unimportant there and is therefore not emphasized. In hysteretic materials, by contrast, handling this distinction is often left to be learned through ``lab know-how'' or brief descriptions in some paper's methods \cite{czajka2021oscillations,singha2023colossal}. The issue has become more pressing with the increasing number of reports of exchange bias and other non-centered hysteresis loops in bulk single-crystalline conductors \cite{lachman2020exchange,noah2022tunable,maniv2021exchange,xu2022ferromagnetic,firdosh2025exchange,kotegawa2023large}, where the methods in Refs.~\cite{czajka2021oscillations,singha2023colossal} do not work.

In this Perspective, we provide a practical reference for processing Hall effect data in magnetically hysteretic materials. We review two specific methods that can be used to mitigate artifacts: (1) the \gls{RMFR} method \cite{sample1987reverse} and (2) antisymmetrization with respect to applied magnetic field between \gls{TRS} which will be referred to as antisymmetrization throughout the manuscript for simplicity.  These methods are general and can also be applied to non-magnetic materials. The \gls{RMFR} method follows directly from the Onsager form of the magnetotransport tensors and relies on interchanging current and voltage contacts. Although it was originally developed for non-magnetic materials \cite{sample1987reverse}, we show here that it remains valid and is particularly useful in hysteretic magnetic materials. Antisymmetrization, by contrast, relies on reversing the applied magnetic field and is more commonly used, but it is also more easily misapplied, which can lead to artifacts that mimic the \gls{AHE} or \gls{THE}. To help identify and mitigate such artifacts, we provide a generic decision tree for Hall measurements in hysteretic materials. We demonstrate and compare both methods using Co$_3$Sn$_2$S$_2$, which can be prepared either with or without \gls{EB}. We further show that both methods remain valid even when the magnetic hysteresis loops are not centered about zero applied field, provided that the data are processed between \gls{TRS}. We also show that \gls{RMFR} can in some cases be twice as time-efficient as antisymmetrization and can be less susceptible to trapped-flux artifacts.

Furthermore, we use measurements on CeCoGe$_3$ \cite{pecharsky1993unusual,thamizhavel2005unique}, a system that can be prepared with asymmetric magnetization and resistivity via specific field-cooling procedures \cite{moya2026tunable}, to demonstrate how incorrect antisymmetrization can produce artifacts that mimic the spontaneous \gls{AHE}. We conclude by discussing how to identify and correct artifacts in Hall data, outlining practical considerations for contact geometry, and highlight classes of materials where these methods will continue to be be especially valuable.

\section{A Detailed View of the Problem}\label{details}
Consider a standard Hall bar geometry within a Cartesian coordinate system, where a current $I_x$ is applied and the transverse voltage $V_y$ is measured in the presence of a magnetic induction field $B_z = \mu_0(H_z + M_z)$, with $H_z$ the applied magnetic field and $M_z$ the magnetization. Within the linear-response regime, the transverse voltage is related to the resistance tensor $R_{ij}(B_z)$ by $V_y(B_z) = I_x R_{yx}(B_z)$.  $R_{yx}$ generally contains two components: an antisymmetric (odd) Hall resistance, $R_{yx}^{odd}(+B_z)$ = -$R_{yx}^{odd}(-B_z)$, and a symmetric (even) transverse magnetoresistance  $R_{yx}^{even}(+B_z)=R_{yx}^{even}(-B_z)$. ``Antisymmetric'' or ``symmetric'' refer to the form of the matrix representing the magnetoresistance, while ``odd'' or ``even'' refer to the specific matrix element's response to the inversion of the magnetic induction field.  Symmetry constrains $R_{yx}$ to be purely odd-in field if the measurement plane has rotational symmetry of order $>2$, or if $B_z$ is perpendicular to an axis with even $n$-fold rotational symmetry and the measurement plane is aligned with that axis \cite{kao1958phenomenological}. Therefore, trigonal, monoclinic, and triclinic crystal systems can have even-in-field $R_{yx}^{even}$ components even when the measurement is done along a high-symmetry orientation. However, the effect can be present in higher-symmetry systems if the current, voltage and magnetic field are not all orthogonal or along principal axes depending on the point group. The point groups that allow even-in-field contributions to $R_{yx}$ are listed by Akg\"{o}z and Saunders \cite{akgoz1975space}, and may also be checked via the magnetoresistance tensor on the Bilbao Crystallographic Server \cite{gallego2019automatic}.

In practice, particularly for bulk single crystals where electrical contacts are applied manually, a voltage contact misalignment introduces an additional longitudinal voltage component $V_x = I_x R_{xx}^{even}(B_z)$ into the measured signal. The measured voltage is therefore a combination, $V_m(B_z) = I_x \left[ R_{yx}^{odd}(B_z) +R_{yx}^{even}(B_z)+  R_{xx}^{even}(B_z) \right]$. To extract the Hall voltage, one typically exploits the Onsager--Casimir reciprocity relations which imply that the antisymmetric components of the intrinsic magnetotransport tensors, i.e., the magnetoconductivity $\boldsymbol{\sigma}$ and the magnetoresistivity $\boldsymbol{\rho}$, are strictly $\mathbf{B}$-odd, and the symmetric components are strictly $\mathbf{B}$-even \cite{onsager1931reciprocal,onsager1931reciprocal2,casimir1945onsager,akgoz1975space}.  Under the usual assumptions of a well-defined current path and measurement geometry, $\rho_{yx} \propto R_{yx}$ and $\rho_{xx} \propto R_{xx}$ up to geometric factors, and therefore the measured resistances inherit these symmetries. Explicitly, $R_{ij}^{even}(+B_z) = R_{ij}^{even}(-B_z)$ and $R_{ij}^{odd}(+B_z) = -R_{ij}^{odd}(-B_z)$. These relations hold for linear, microscopically reversible systems. By subtracting (antisymmetrizing) the measured voltage under $B_z$ reversal, one cancels the $B_z$-even components, yielding the antisymmetric component of the tensor,

\begin{equation}
     R_{yx}^{odd}(B_z)  = \frac{R_m(+B_z) - R_m(-B_z)}{2},
    \label{Eq:paramagnetic}
\end{equation}

\noindent where $R_m = V_m/I_x$.

A conundrum arises because in an electrical transport measurement the following aspects are true: only $H_z$ is controlled, $M_z$ can be path dependent, and is it not directly measured, while Onsager--Casimir reciprocity only holds for $B_z = \mu_0(H_z+M_z)$. In materials without magnetic hysteresis, the conundrum is easily avoided, since $M_z$, and consequently $B_z$, are antisymmetric with respect to $H~=~0$. Thus, the Onsager--Casimir relations also hold for $H_z$ and

\begin{equation}
     R_{yx}^{odd}(H_z)  = \frac{R_m(+H_z) - R_m(-H_z)}{2}.
    \label{Eq:Hall(H)}
\end{equation}

However, in materials with magnetic hysteresis-including ferromagnets, antiferromagnets with complicated domains, metamagnets, glassy magnets and \gls{EB} systems (hysteresis loops shifted away from $H = 0$) - $M(H)$ and $R_{ij}(H)$ may be path dependent and multi-valued. In cases of magnetic hysteresis, the Onsager--Casimir reciprocity relations can still hold in principle, but only between fully \gls{TRS}. Experimentally, this means that to use the Onsager--Casimir relations, one must prepare the sample with time-reversed $B_z$, which may not correspond to equal and opposite $H_z$ for a continuous sweep. This mismatch between the magnetic induction field and applied magnetic field is the origin of the artifacts discussed in this Perspective.

\begin{figure}[htbp]
 \centering
\includegraphics[width = 0.8\columnwidth]{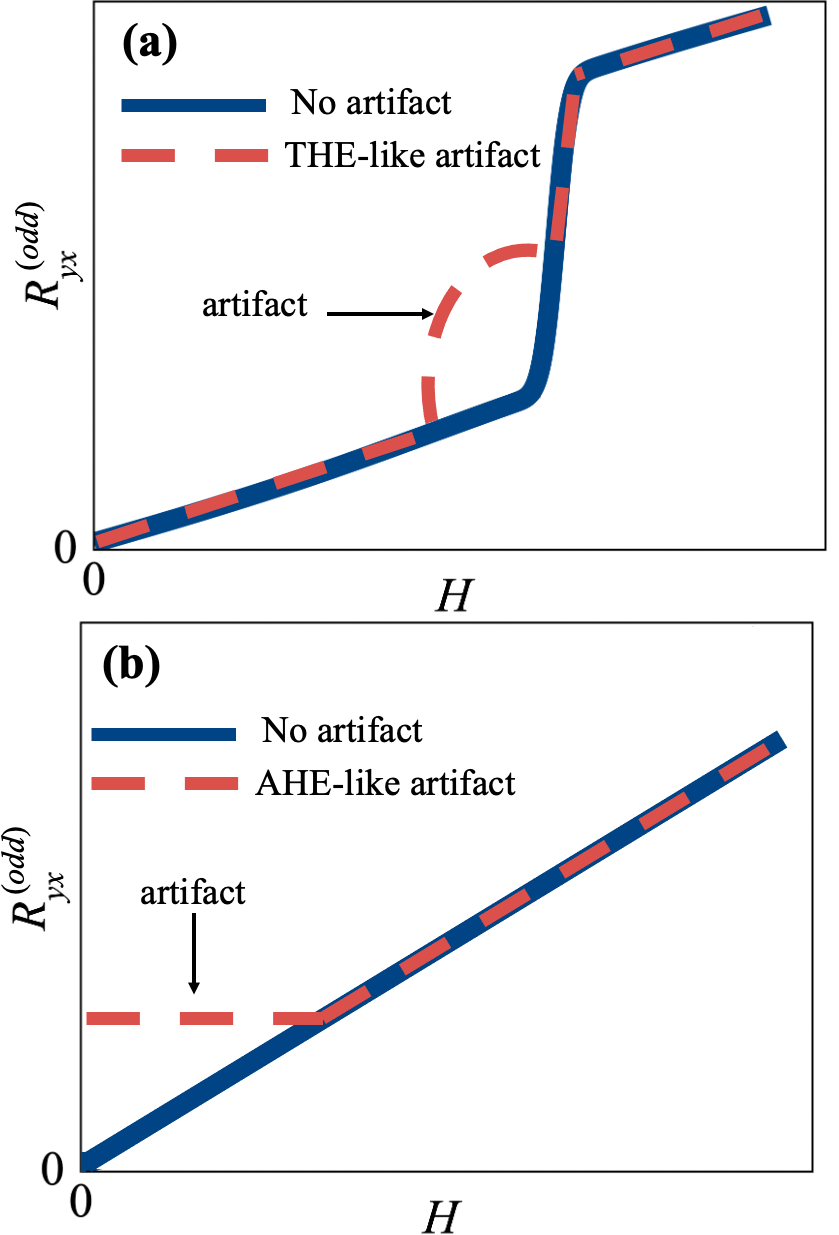}
\caption{\label{cartoon}
Schematics of (\textbf{a}) topological Hall effect \gls{THE}-like and (\textbf{b}) anomalous Hall effect \gls{AHE} - like artifacts that can arise in measuring the magnetic field $H$ dependent Hall resistance $R_{yx}^{odd}$. The blue lines symbolize data with no artifact, and the red dashed lines symbolize the signal contaminated with experimental artifacts. The (\textbf{a}) \gls{THE}-like artifact shows up as an unexpected feature at finite applied magnetic field $H$, while the (\textbf{b}) \gls{AHE}-like artifact gives finite  $R_{yx}^{odd}$ at $H$= 0. 
}
\end{figure}

Schematics of common \gls{THE}- and \gls{AHE}-like artifacts are shown in Figure~\ref{cartoon}a,b, respectively. These artifacts can arise either from improper treatment of hysteresis or from trapped flux in a superconducting magnet. Before discussing how these artifacts can be mitigated, we first introduce Co$_3$Sn$_2$S$_2$, which we use as an example material to demonstrate the \gls{RMFR} and antisymmetrization methods.

%The main advantages of the \gls{RMFR} method are threefold: (1) for a given $H$--$T$ history it yields twice as much independent Hall data and is therefore twice as time-efficient as antisymmetrization, (2) it mitigates artifacts from trapped flux, and (3) its procedure is independent of whether hysteresis is present. In contrast, antisymmetrization is useful when only one source/measure channel is available, whereas \gls{RMFR} requires either two source/measure channels or a switch box. However, as we will demonstrate, the antisymmetrization procedure depends on the type of hysteresis and can suffer from artifacts due to trapped flux.

%We also use measurements on CeCoGe$_3$ \cite{pecharsky1993unusual,thamizhavel2005unique}, a system that can be prepared with asymmetric magnetization and resistivity via specific field-cooling procedures \cite{moya2026tunable}, to demonstrate how incorrect antisymmetrization can produce artifacts that mimic the spontaneous \gls{AHE}. We conclude by discussing how to identify and correct artifacts in Hall data, outlining practical considerations for contact geometry, and highlighting a class of materials where these methods should be especially valuable.

\section{The Test Case: C\MakeLowercase{o}$_3$S\MakeLowercase{n}$_2$S$_2$}

% Figure 2 is separated from its full caption so that the caption can occupy its own page.
\clearpage
\begin{figure}[p]
 \centering
 \refstepcounter{figure}\label{Co3HallM}
 \includegraphics[width = 0.8\textwidth]{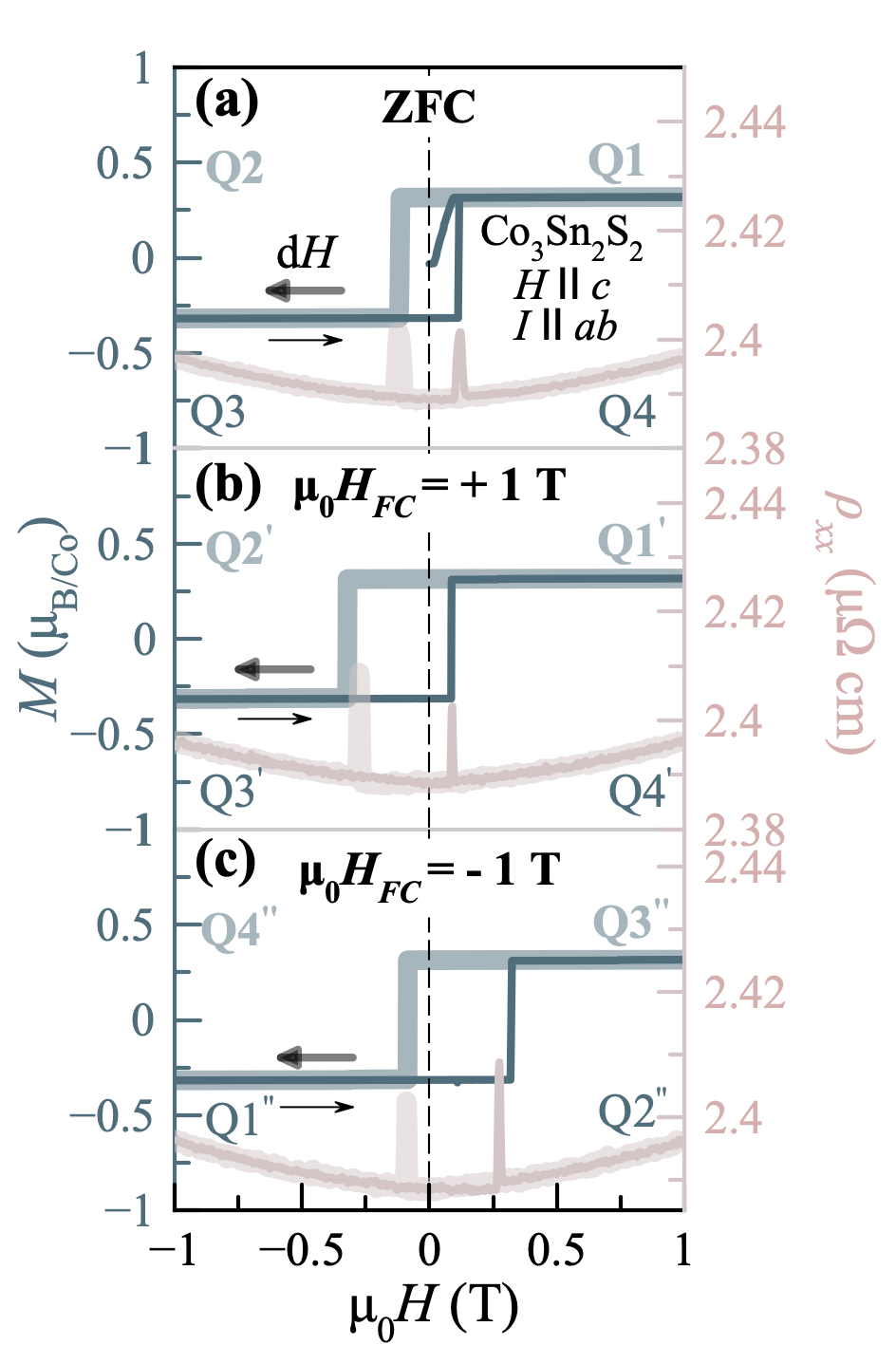}
 \vspace{0.5em}
 
 \noindent\textbf{Figure~\thefigure.} Full caption on following page.
\end{figure}
\clearpage
\thispagestyle{plain}
\begingroup
\small
\noindent\textbf{Figure~\ref{Co3HallM}.} Isothermal magnetization ($M$) (blue, left axis) and longitudinal resistivity ($\rho_{xx}$) (pink, right axis) measured on Co$_3$Sn$_2$S$_2$ at $T$ = 5 K after cooling the sample from 300 K using the (\textbf{a}) zero-field cooled (\gls{ZFC}) protocol, (\textbf{b})  positive field cool protocol with $\mu_0H_{FC}~=~+1$~T and (\textbf{c}) negative field cool protocol with $\mu_0H_{FC}~=~-1$~T. 
For the \gls{ZFC} protocol in (\textbf{a}), the sample was cooled in zero field from $T_{max}=300$ K to $T_{meas}=5$ K, then the magnetic field was swept $0 \rightarrow +1~\text{T} \rightarrow -1~\text{T} \rightarrow +1~\text{T}$. 
For the positive field-cool protocol in (\textbf{b}), the sample was cooled from $T_{max}$ to $T_{meas}$ in $\mu_0 H_{FC}=+1$ T, and the field was swept $+1~\text{T} \rightarrow -1~\text{T} \rightarrow +1~\text{T}$.  For the negative field-cool protocol in (\textbf{c}), the sample was cooled from $T_{max}$ to $T_{meas}$ in $\mu_0 H_{FC}=-1$ T, and the field was swept $-1~\text{T} \rightarrow +1~\text{T} \rightarrow -1~\text{T}$.  Q1-Q4 label the sequential quarter-segments of the field sweep; for the (\textbf{a}) \gls{ZFC} loop these correspond to $+1~\text{T}\rightarrow 0$, $0\rightarrow -1~\text{T}$, $-1~\text{T}\rightarrow 0$, and $0\rightarrow +1~\text{T}$, respectively. Q1$'$-Q4$'$ denote the analogous segments for the (\textbf{b}) positive  field-cool protocol, while Q1$''$-Q4$''$ in (\textbf{c}) sequentially correspond to $-1~\text{T}\rightarrow 0$, $0\rightarrow 1~\text{T}$, $1~\text{T}\rightarrow 0$, and $0\rightarrow -1~\text{T}$ (\textbf{c}) for the negative field-cool protocol.   All measurements were performed with the applied magnetic field $H \parallel c$ while the transport measurements were performed with the current $I\parallel ab$.
\par
\endgroup
\clearpage

Co$_3$Sn$_2$S$_2$, a Shandite compound crystallizing in the $R3\bar m$ (No. 166) space group that orders ferromagnetically near $T_C ~\sim$ 175 K \cite{weihrich2006half,vaqueiro2009powder,kubodera2006ni,schnelle2013ferromagnetic}, is an ideal platform to test measuring the \gls{AHE} in cases where the magnetic hysteresis is or is not centered about $H~=~0$ . With application of $H\parallel c$, a large \gls{AHE} is registered below $T_C$ which was linked to Weyl fermions in its Fermi surface \cite{liu2018giant,wang2018large}. A second anomaly was observed in various measurements near $T_A~\sim~125$ K \cite{kassem2017low,lachman2020exchange,okamura2020giant,lee2022observation,shen2022anomalous} with its origins debated \cite{guguchia2020tunable,lee2022observation,soh2022magnetic,zhang2022hidden,menil2025magnetic}. Below $T_A$, magnetic memory effects are observed \cite{menil2025magnetic}, including an \gls{EB} \gls{AHE} which can be achieved by field cooling the sample from above $T_C$ through $T_A$ likely due to magnetic phase coexistence \cite{lachman2020exchange,noah2022tunable}. 

In the original report of the \gls{EB} \gls{AHE} in Co$_3$Sn$_2$S$_2$, the extrinsic quantity, the Hall \textit{resistance}, was reported where the longitudinal component of the measured resistance was subtracted in such a way that did not allow for reporting of the intrinsic quantity, the Hall \textit{resistivity},  due to the fact that the magnetic hysteresis is by definition not centered around $H~=~0$ \cite{lachman2020exchange}. In determining the origins of \gls{AHE}, the intrinsic magnitude of the Hall conductivity (and thereby the Hall resistivity) is an essential quantity since it can be compared directly with theoretical calculations. Furthermore, the methods used in that report take advantage of the fact that the Hall voltage is exceptionally large in Co$_3$Sn$_2$S$_2$, which is not generally true for most magnetic materials. This further creates a clear impetus to establish generic procedures for extracting the Hall response in systems that show non-centered hysteresis loops.

We have grown single crystals of Co$_3$Sn$_2$S$_2$ out of a ternary melt described in Methods. The crystal structure was checked using X-ray diffraction (Figure S1, Tables S1--S4) and refinements are consistent with previous reports. In Figure~\ref{Co3HallM}a--c we present $H$-dependent isothermal magnetization $M$ (blue, left axis) and longitudinal resistivity $\rho_{xx}$ (pink, right axis) measured on Co$_3$Sn$_2$S$_2$ at temperature $T~=~5$ K with $H\parallel c$ under various field cooling procedures. %The corresponding Hall resistivity $\rho_{yx}^{odd}$ is plotted in Figure~\ref{Co3HallM}d-e. 

The field cooling procedures are shown schematically in Figure~\ref{summaryfig}a,b. Explicitly, for the \gls{ZFC} case,  the crystal was cooled in the absence of $H$ from $T_{max}~=~300$ K to $T_{meas}~=~5$ K. $H$ was then ramped from $0\rightarrow+1~\text{T}\rightarrow-1~\text{T}\rightarrow1~\text{T}$. Here, no \gls{EB} behavior is observed as the hysteresis loops are centered around $H~=~0$ in $M$ and $\rho_{xx}$ (Figure~\ref{Co3HallM}a). To induce the \gls{EB} effect, the sample was field cooled at $\mu_0H_{FC}~=~+1$ T from 300 K then a hysteresis loop was measured at $T~=~5$~K from $+1~\text{T}\rightarrow-1~\text{T}\rightarrow1~\text{T}$, schematically shown in Figure~\ref{summaryfig}b as the positive FC procedure. The corresponding procedure for $\mu_0H_{FC}~=~-1$ T is labeled as the negative FC procedure in Figure~\ref{summaryfig}b where the field cooling procedure was the same but the field cooling and field sweep protocol was reversed i.e the sample was field cooled in $\mu_0H_{FC}~=~-1$ T from 300 K and the hysteresis loops were measured from $-1~\text{T}\rightarrow1~\text{T}\rightarrow-1~\text{T}$. The \gls{EB} effect  is observed as a bias of the $M$ and $\rho_{xx}$ hysteresis loops towards the left of $H~=~0$ for $\mu_0H_{FC}~=~+1$ T (Figure~\ref{Co3HallM}b)  and to the right for $\mu_0H_{FC}~=~-1$ T (Figure~\ref{Co3HallM}c)  consistent with previous measurements \cite{lachman2020exchange}.  In all three of these cases there are regions where $M(H) \neq -M(-H)$, and therefore Equation~\ref{Eq:Hall(H)} cannot be used to antisymmetrize the Hall data. In the following, we describe two methods to extract the Hall effect and demonstrate them on Co$_3$Sn$_2$S$_2$.

\section{Methods for Extracting the Hall Effect}\label{RMFRvAnti}

% Figure 3 is separated from its full caption so that the caption can occupy its own page.
\clearpage
\begin{figure}[p]
\centering
\refstepcounter{figure}\label{summaryfig}
\includegraphics[width=0.7\textwidth]{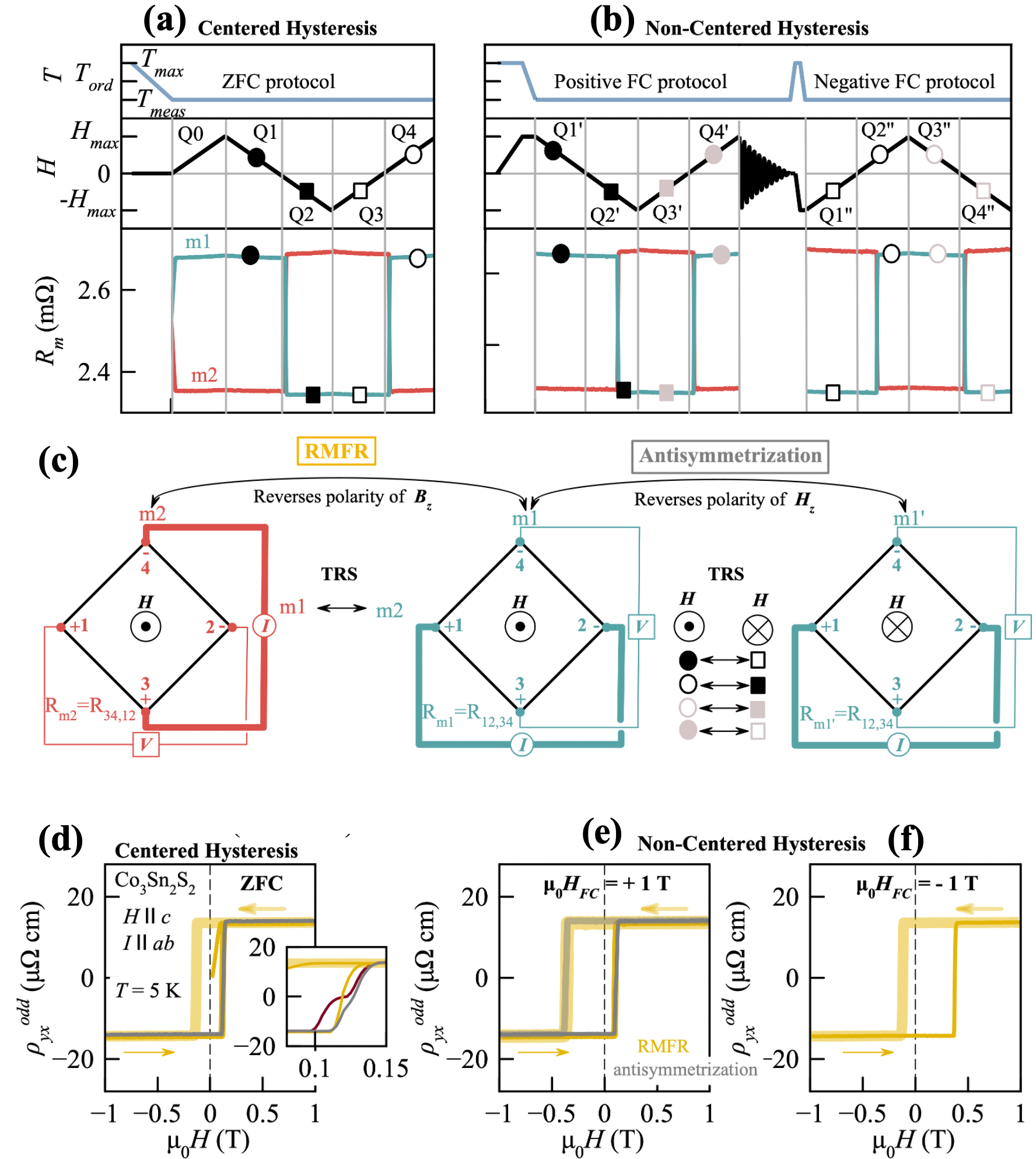}
\vspace{0.5em}

\noindent\textbf{Figure~\thefigure.} Full caption on following page.
\end{figure}
\clearpage
\thispagestyle{plain}
\begingroup
\small
\noindent\textbf{Figure~\ref{summaryfig}.} Generic time ($t$) histories of magnetic field ($H$) and temperature ($T$) for (\textbf{a}, top axis) the zero-field-cooled (\gls{ZFC}) protocol and (\textbf{b}, top axis) the positive and negative field-cooled (FC) protocols. For Co$_3$Sn$_2$S$_2$ with applied field $H \parallel c$, $\mu_0 H_{max}=1$ T, $T_{max}=300$ K, and $T_{meas}=5$ K, the \gls{ZFC} protocol yields centered hysteresis loops whereas FC protocols yield non-centered loops. Q labels indicate the measurement order of the field-sweep segments: for \gls{ZFC}, Q0 denotes $0 \rightarrow +1~\text{T}$, followed by Q1--Q4: $+1~\text{T} \rightarrow 0$, $0 \rightarrow -1~\text{T}$, $-1~\text{T} \rightarrow 0$, and $0 \rightarrow +1~\text{T}$. For the positive FC protocol, Q1$'$--Q4$'$ use the same segment definitions as Q1--Q4 but without Q0. For the negative FC protocol, Q1$''$--Q4$''$ sequentially label $-1~\text{T} \rightarrow 0$, $0 \rightarrow +1~\text{T}$, $+1~\text{T} \rightarrow 0$, and $0 \rightarrow -1~\text{T}$. Corresponding histories of the measured resistances $R_m$, labeled m1 (blue) and m2 (red), used to extract the Hall resistivity $\rho_{yx}^{odd}$ of Co$_3$Sn$_2$S$_2$ are shown for (\textbf{a}, bottom axis) the \gls{ZFC} protocol and (\textbf{b}, bottom axis) the positive and negative FC protocols. The measurements were performed with current $I \parallel ab$. Shown for a Van der Pauw geometry (\textbf{c}), m1 and m2 correspond to $R_{m1}=R_{12,34}$ and $R_{m2}=R_{34,12}$, where the first index pair denotes the current source and drain contacts and the second index pair denotes the high and low voltage contacts. (\textbf{c}) Schematic comparison of reverse-magnetic-field reciprocity (\gls{RMFR}) and antisymmetrization methods for extracting the Hall effect. The \gls{RMFR} method uses m1 and the $I$--$V$-rotated measurement m2 to generate a pair of measurements equivalent to reversing the polarity of $B_z$ resulting in time-reversed states (TRS), while antisymmetrization uses m1 and m1$'$ measured with the same contact geometry at equal and opposite $H$ to measure TRS. Depending on whether the hysteresis loops are centered or non-centered, one or two hysteresis loops are required to access $H$-dependent TRS; the \gls{TRS} pairing used in (\textbf{a,b}) is encoded according to the legend in (\textbf{c}). $\rho_{yx}^{odd}$ obtained from $R_m$ in (\textbf{a,b}) for (\textbf{d}) the \gls{ZFC} protocol, (\textbf{e}) the positive FC protocol with $\mu_0 H_{FC}=+1$ T, and (\textbf{f}) the negative FC protocol with $\mu_0 H_{FC}=-1$ T. Yellow curves in (\textbf{d--f}) are obtained using \gls{RMFR}, while gray curves are obtained using antisymmetrization. Thick lines denote sweeps from $+1$ T$\rightarrow -1$ T, and thin lines denote $-1$ T$\rightarrow +1$ T. The inset of (\textbf{e}) highlights the coercive-field region; the maroon data are an independent measurement using the antisymmetrization method.
\par
\endgroup
\clearpage

We first introduce the \gls{RMFR} and antisymmetrization methods. Schematics of two measurements, labeled m1 (blue) and m2 (red) are shown in Figure~\ref{summaryfig}c, where the current and voltage contacts are interchanged relative to each other. Physically, the \gls{RMFR}  method \cite{sample1987reverse} follows from Onsager--Casimir reciprocity of the magnetotransport tensors. For a linear, ohmic conductor this implies that in a four-terminal resistance device, the voltage pattern produced by driving current from contact $1$ to $2$ in field $+B_z$ (Figure~\ref{summaryfig}c (m1, blue)) is identical to that produced by driving the same current from $3$ to $4$ in $-B_z$ (Figure~\ref{summaryfig}c (m2, red)) with the roles of current and voltage contacts interchanged. Thus, interchanging current and voltage leads at fixed $B_z$ is equivalent to keeping
the leads fixed and reversing the magnetic induction field ($+B_z\rightarrow-B_z$). Therefore, measurements m1(blue) and m2(red) are \gls{TRS} equivalents. Since rotating the electrical contacts equivalently reverses the polarity of $B$ for any $H$,  Equation~\ref{Eq:Hall(H)} can be  re-written as 

\begin{equation}
R_{yx}^{odd}(H_z) = \frac{R_{m1}(H_z)-R_{m2}(H_z)}{2}.
    \label{Eq:spinningcurrents}
\end{equation}

\noindent This procedure assumes that appropriate averaging techniques are used to eliminate thermoelectric voltages. In a \gls{DC} transport measurement, this can be achieved by reversing the current and appropriately averaging such that $R_{m1}~=~\frac{R_{12,34}-R_{21,34}}{2}$ and $R_{m2}~=~\frac{R_{34,12}-R_{43,12}}{2}$, where the first indices represent the current source and drain contacts, and the second indices represent the high and low limit voltage contacts \cite{lindemuth2020hall}. 

Instead, we adopt the common practice of using low-frequency \gls{AC} lock-in transport measurements to eliminate thermoelectric voltages and approximate the \gls{DC} limit reducing the procedure to two measurements with $R_{m1}~=~R_{12,34}$ and $R_{m2}~=~R_{34,12}$. Furthermore, the \gls{AC} lock-in technique allows for simultaneous acquisition of $R_{m1}$ and $R_{m2}$ if two \gls{AC} current sources and voltage measure modules are available and appropriate lock-in and grounding techniques are considered. The techniques we used for simultaneous acquisition of $R_{m1}$ and $R_{m2}$ are described in Methods.

In general, the only constraints of the \gls{RMFR} method are that the device under test be ohmic (the current--voltage relationship is linear, which we have checked in our devices), and the states at $+B_z$ and $-B_z$ are time-reversed counterparts (i.e., the Onsager--Casimir relations hold) \cite{sample1987reverse}. The validity of the Onsager--Casimir relations already presumes linear response, but we state the ohmic requirement explicitly because, in experiment, non-linearity can arise from extrinsic effects such as Joule heating or contact-related artifacts, as well as from intrinsic non-linear or non-reciprocal transport mechanisms. Accordingly, RMFR should not be expected to hold once the measurement current drives the sample out of linear response, or when Onsager--Casimir reciprocity is broken intrinsically. While we have shown m1 (blue) and m2 (red) in Figure~\ref{summaryfig}c for a \gls{VdP} geometry, Hall bar or other more arbitrary geometries can also work for the \gls{RMFR} method \cite{cornils2008reverse}. It goes without saying that the same exact contacts should be used for both measurements. So long as these conditions are met and $R_{m1}$ and $R_{m2}$ are taken simultaneously or sequentially (if simultaneous acquisition is not possible) without changing $H$, the \gls{RMFR} method together with Equation~\ref{Eq:spinningcurrents} can be used to extract the Hall effect for an arbitrary $H-T$ history.

This is in stark contrast to isolating Hall signals by antisymmetrizing with respect to $H$. Here we will refer to this process as ``antisymmetrization.'' The idea is that two measurements m1 and m1' are taken with the same contact geometry at equal and opposite $H$, as shown schematically in Figure~\ref{summaryfig}c for a \gls{VdP} geometry. Again, antisymmetrization also holds for Hall bar or more arbitrary geometries. Like the \gls{RMFR} method, antisymmetrization also assumes ohmic devices and appropriate averaging techniques are used to eliminate thermoelectric voltages, where for our measurements we opt for the low-frequency \gls{AC} lock-in method.  

Although $M(H_z)\neq-M(-H_z)$ for all $H_z$ in hysteretic materials, the experimental variable $H_z$ can still be used to antisymmetrize the measured signal, but special care must be taken so that the system is prepared such that the Onsager--Casimir relations hold for $H_z$. In general, there are two possible scenarios: (1) where magnetic hysteresis loops are centered about $H~=~0$ (the \gls{ZFC} case of Co$_3$Sn$_2$S$_2$ [Figure~\ref{Co3HallM}a]) which we will refer to as centered hysteresis loops and (2) where the magnetic hysteresis is not centered about $H~=~0$ (the \gls{EB} case of Co$_3$Sn$_2$S$_2$  (Figure~\ref{Co3HallM}b,c) which we will refer to as non-centered hysteresis loops.

\begin{figure}[htbp]
\centering
\includegraphics[width=1\textwidth]{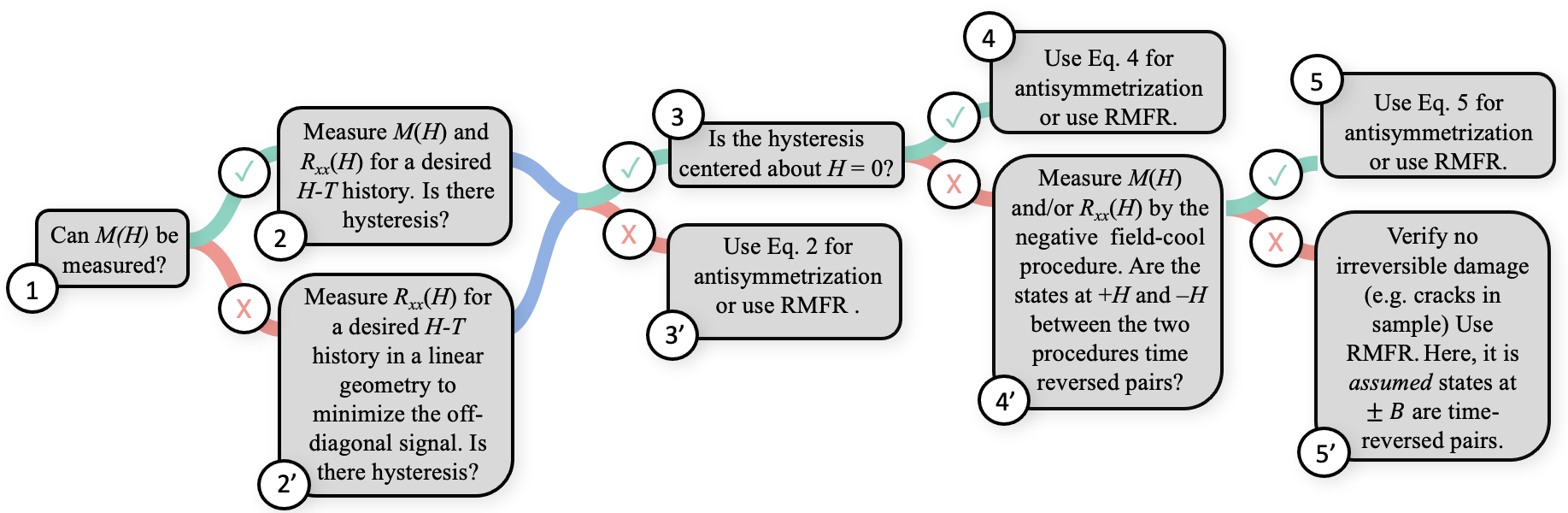}
\caption{\label{flowchart} A decision tree that can be used to minimize experimental artifacts in Hall effect measurements. } 
\end{figure}

With the two methods introduced, we demonstrate how to extract the Hall effect shown in Figure~\ref{summaryfig}d, for the centered hysteresis loop, and Figure~\ref{summaryfig}e,f for the \gls{EB} state of Co$_3$Sn$_2$S$_2$  using the decision tree presented in Figure~\ref{flowchart}.

\subsection{Case 1: Centered Hysteresis Loops}\label{sec:centered}

Figure~\ref{flowchart} presents a generic flow chart for the steps in measuring the Hall effect, where we first recommend measuring $M(H)$ (Box 1) for the same $H$ orientation and history $H-T$ for which it is  desired to measure the Hall effect. Since the Hall effect cannot be measured directly without perfectly aligned contacts, $M(H)$ is convenient method of checking for hysteresis as it can be measured without any data manipulation and the Hall effect directly depends on it through $B$. If it is not possible to measure $M(H)$ on your sample, measuring $R_{xx}(H)$ in a collinear geometry where all contacts span the entire width of the sample to minimize the parasitic Hall voltages, is the next best thing to check for hysteresis (Path 1-2'). Either way, measuring $R_{xx}$ even if $M(H)$ can also be measured (Path 1-2) is useful for converting the transport measurements to conductivity, but can also be used to diagnose artifacts as will be demonstrated later. If there is no hysteresis, $R_{yx}^{odd}(H)$ can be measured (1) with the \gls{RMFR} method in a single field sweep from $0\rightarrow H_{max}$ using Equation~\ref{Eq:spinningcurrents} or (2) with antisymmetrization method sweeping from $-H_{max}\rightarrow H_{max}$ using Equation~\ref{Eq:Hall(H)}. Either way, the $H-T$ path of the measurement should not matter since $M(+H)~=~M(-H)$ for all $H$. This is not the case in hysteretic materials, as will now be demonstrated.

If \gls{EB} or other strong history dependence is suspected, it is also useful to compare \gls{ZFC} and positive and negative field-cooled protocols. Diagnostic signatures include when features in $M(H)$ or $R_{xx}(H)$  hysteresis loops shift away from $H=0$. Such checks are not necessary in every material, but they are a practical way to determine whether a nominally centered \gls{ZFC} loop is masking a field-history-dependent state which can give knowledge about the pinning in the material of interest.

Following Path 1-2-3-4 in Figure~\ref{flowchart} we return to the \gls{ZFC} loop of $M$ and $\rho_{xx}$ in Figure~\ref{Co3HallM}a, where the $H-T$ history of these measurements is plotted in Figure~\ref{summaryfig}a. The measurements can be separated into five quadrants: \textit{Q}0 ($0 \rightarrow H_{max}$), \textit{Q}1 ($H_{max}\rightarrow0$), \textit{Q}2 ($0\rightarrow -H_{max}$), \textit{Q}3 ($-H_{max}\rightarrow0$), and \textit{Q}4 ($0\rightarrow H_{max}$). Focusing on $M$ in Figure~\ref{Co3HallM}a, it is apparent that $M(Q1)=-M(Q3)$ and $M(Q4)=-M(Q2)$, that is, these respective quadrants are \gls{TRS} of each other. Note, that no \gls{TRS} of \textit{Q}0 is present for such a \gls{ZFC} hysteresis loop in $M$.
Thus, to obtain $R_{yx}^{odd}$ using antisymmetrization, $R_{m1}$, shown in Figure \ref{summaryfig}a, (blue), must be antisymmetrized in a piecewise manner using

\begin{equation}
R_{yx}^{odd}(H) = \begin{cases}
\frac{R_{m1}(Q1,~\fcirc{black})-R_{m1}(Q3,~\fsq{white})}{2} & \text{for}~[Q1] \\
\frac{R_{m1}(Q2,~\fsq{black})-R_{m1}(Q4,~\fcirc{white})}{2} & \text{for}~[Q2] \\
\frac{R_{m1}(Q3,~\fsq{white})-R_{m1}(Q1,~\fcirc{black})}{2} & \text{for}~[Q3]\\
\frac{R_{m1}(Q4,~\fcirc{white})-R_{m1}(Q2~\fsq{black})}{2} & \text{for}~[Q4]. \\

\end{cases}
\label{Eq:ZFC}
\end{equation}

\noindent In other words, Equation~\ref{Eq:ZFC} is applied piecewise within a single centered hysteresis loop by antisymmetrizing only those sweep segments that are time-reversed counterparts. For the case shown here, this means pairing $Q1$ with $Q3$ and $Q4$ with $Q2$. This is why a full hysteresis loop is required even when the loop is centered about $H = 0$.

Analyzing Equation~\ref{Eq:ZFC}, it is apparent that such a five quadrant hysteresis loop only produces two quadrants of \textit{independent} data, either $R_{yx}^{odd}(Q1\rightarrow Q2)$ or $R_{yx}^{odd}(Q3\rightarrow Q4)$. By symmetry $R_{yx}^{odd}(Q1\rightarrow Q2)$ =  -$R_{yx}^{odd}(Q3\rightarrow Q4)$. While it is common practice to report all four quadrants, to emphasize the fact that only two quadrants of independent Hall data can be obtained by such a measurement of $R_m$, we only plot $\rho_{yx}^{odd}(Q3\rightarrow Q4)$ in Figure~\ref{summaryfig}d (gray) obtained by antisymmetrizing  $R_{m1}$ (Figure \ref{summaryfig}a, blue) using Equation~\ref{Eq:ZFC} and multiplying by the thickness of the sample. If it is desired to extract $R_{yx}^{odd}$ of the virgin \gls{ZFC} branch using antisymmetrization, an additional \gls{ZFC} measurement must be performed with the initial field sweep reversed, i.e. $0\rightarrow -H_{max}$, so that the virgin branch measured for $0\rightarrow H_{max}$ can be paired with its time-reversed counterpart.

By contrast, the \gls{RMFR} method records $R_{m1}$ (Figure~\ref{summaryfig}a, blue) and $R_{m2}$ (Figure~\ref{summaryfig}a, red) simultaneously along the same \gls{ZFC} $H-T$ history, allowing $\rho_{yx}^{odd}$, to be obtained via Equation~\ref{Eq:spinningcurrents}. The corresponding  $\rho_{yx}^{odd}$ is presented in Figure~\ref{summaryfig}d (yellow). Notably, \textit{independent} measurements of all five measurement quadrants are obtained. Mirroring $M$ and $\rho_{xx}$ (Figure~\ref{Co3HallM}a), $\rho_{yx}^{odd}$ (Figure~\ref{summaryfig}d) exhibits no exchange bias in the \gls{ZFC} case of Co$_3$Sn$_2$S$_2$.

Overall, the \gls{RMFR} method (yellow) and the antisymmetrization method (gray) agree within experimental accuracy on both the spontaneous ($H=0$) \gls{AHE} and the high-field value of $\rho_{yx}^{odd}$. However, differences emerge near the coercive fields (Figure~\ref{summaryfig}d, inset): the \gls{RMFR} yields a smooth sign reversal of $\rho_{yx}^{odd}$, whereas antisymmetrization shows a kink, and the coercive field obtained by the two methods is slightly offset. An independent Hall-bar measurement processed with the same antisymmetrization procedure (Equation~\ref{Eq:ZFC}) accentuates these discrepancies (maroon curve in Figure~\ref{summaryfig}d).

These deviations are naturally explained by a small trapped field in the superconducting magnet. Because antisymmetrization assumes perfectly opposite fields, the slight mismatch in switching fields where $\rho_{xx}$ changes rapidly (Figure~\ref{Co3HallM}a) produces the observed kink. By construction, the \gls{RMFR} pairs measurements that simulate an exactly reversed $B_z$ at the same $H_z$ (Figure~\ref{summaryfig}c), mitigating this specific trapped-flux artifact. More broadly, anomalies around the coercive fields warrant caution as we will elaborate more on in the discussion. As for mitigating the coercive field issue, a finite offset in the magnitude of $H_z$ is difficult to completely eliminate when using a superconducting magnet. It can be reduced by following the manufacturer degaussing/oscillation-to-zero procedures prior to measurement.

\subsection{Case 2: Non-Centered Hysteresis Loops\label{non-centered}}
In the case where there is an \gls{EB} effect or non-centered hysteresis loop in either $M$ or $\rho_{xx}$ (Path 1-2-3-4'-5 in Figure~\ref{flowchart}), as is the case for Co$_3$Sn$_2$S$_2$ in the field-cooled states (Figures~\ref{Co3HallM}b,c), two separate hysteresis loops are needed to extract the Hall effect if antisymmetrization is used: a positive field-cool protocol and negative field-cool protocol. For Co$_3$Sn$_2$S$_2$, the $H-T$ history for the positive and negative field-cool protocols is schematically shown in Figure~\ref{summaryfig}b with $T_{max}=300~\text{K}>T_{ord}=T_c>T_{meas} = 5~\text{K}$ and $\mu_0H_{max} = \mu_0H_{FC}~=~\pm1$ T. Explicitly, for the $\mu_0H_{FC}~=~+1$ T case measurements are performed for \textit{Q}1' ($H_{max}\rightarrow0$), \textit{Q}2' ($0\rightarrow -H_{max}$), \textit{Q}3' ($-H_{max}\rightarrow0$), and \textit{Q}4' ($0\rightarrow H_{max}$), while for the $\mu_0H_{FC}~=~-1$ T case we have \textit{Q}1'' (-$H_{max}\rightarrow0$), \textit{Q}2'' ($0\rightarrow H_{max}$), \textit{Q}3'' ($H_{max}\rightarrow0$), and \textit{Q}4'' ($0\rightarrow -H_{max}$). Here, we note that the ramp rate should always be kept constant between the positive and negative field cool procedures, as the dynamics in glassy systems can change with sweep rate. While most often non-centered hysteresis loops are found by field-cooling procedures, there are cases where \gls{EB} has been reported in the \gls{ZFC} state of systems \cite{wang2011large,bufaiccal2024essential}. In such a case the positive field-cool protocols would include \gls{ZFC}ing the sample from above $T_{ord}$ and running a hysteresis loop from $0\rightarrow H_{max}\rightarrow -H_{max}\rightarrow H_{max}$ while the negative field-cool protocol would entail the loop from $0\rightarrow -H_{max}\rightarrow H_{max}\rightarrow -H_{max}$.

Returning to the \gls{EB} states in $M$ for Co$_3$Sn$_2$S$_2$ (Figure~\ref{Co3HallM}b,c), it is apparent $M(Q1')\neq -M(Q3')$ and $M(Q2')\neq -M(Q4')$ for all $H$. Therefore, the states are not \gls{TRS} and Equation~\ref{Eq:ZFC} \textit{cannot} be used. However, the \gls{TRS} are prepared using the negative field-cooling protocol where it is demonstrated in Figure~\ref{Co3HallM}b,c M(\textit{Q}1') = -M(\textit{Q}1''), M(\textit{Q}2') = -M(\textit{Q}2''), M(\textit{Q}3') = -M(\textit{Q}3''), M(\textit{Q}4') = -M(\textit{Q}4''). Therefore, Onsager--Casimir reciprocity holds between such time-reversed states,  and $R_{yx}^{odd}$ can be extracted from $R_{m1}$:

\begin{equation}
R_{yx}^{odd}(H) = \begin{cases}
\frac{R_m(Q1', \fcirc{black})-R_m(Q1'',\fsq{white})}{2} & \text{for}~H[Q1', \fcirc{black}] \\

\frac{R_m(Q2', \fsq{black})-R_m(Q2'',\fcirc{white})}{2} & \text{for}~H[Q2', \fsq{black}] \\

\frac{R_m(Q3', \sqmark[lavender])-R_m(Q3'',\circmark)}{2} & \text{for}~H[Q3', \sqmark[lavender]] \\

\frac{R_m(Q4', \sqmark)-R_m(Q4'',\circmark[lavender])}{2} & \text{for}~H[Q4', \sqmark] \\

\frac{R_{m1}(Q1'', \fsq{white})-R_m(Q1',\fcirc{black})}{2} & \text{for}~H[Q1'', \fsq{white}] \\

\frac{R_{m1}(Q2'', \fcirc{white})-R_m(Q2',\fsq{black})}{2} & \text{for}~H[Q2'', \fcirc{white}] \\

\frac{R_{m1}(Q3'', \circmark)-R_m(Q3',\sqmark[lavender])}{2}& 
\text{for}~H[Q3'', \sqmark[lavender]] \\

\frac{R_{m1}(Q4'', \circmark[lavender])-R_m(Q4',\sqmark)}{2} & \text{for}~H[Q4'', \circmark[lavender]]. \\
\label{Eq:FC}
\end{cases}
\end{equation}

\noindent Equation~\ref{Eq:FC} differs from Equation~\ref{Eq:ZFC} because, for non-centered hysteresis, the \gls{TRS} do not occur within a single loop. Instead, the antisymmetrization must be carried out between measurements taken after opposite field-cooling procedures, so that each branch is paired only with its true time-reversed counterpart. In practice, this means that two hysteresis loops are required. If the non-centered hysteresis trains (coercive fields change) with repeated field cycling, antisymmetrization should be carried out only between loops with the same cycle index and otherwise identical field--temperature history under positive and negative field-cooling conditions. If such matched \gls{TRS} cannot be reproducibly prepared, the \gls{RMFR} method is preferred as suggested in Path 1-2-3-4-5' of Figure~\ref{flowchart}.

Again we emphasize that even though eight quadrants were measured, only four quadrants of \textit{independent} data are obtained: either $Q1'\rightarrow Q4'$ or $Q1''\rightarrow Q4''$. Hence, we plot $\rho_{yx}^{odd}$ in Figure~\ref{summaryfig}e (gray) only for $Q1'\rightarrow Q4'$ extracted from $R_{m1}$ in Figure~\ref{summaryfig}b using antisymmetrization by Equation~\ref{Eq:FC}.

Using the \gls{RMFR} method by measuring m1 (blue) and m2 (red) simultaneously for the same field-cool protocols in Figure~\ref{summaryfig}b, two independent hysteresis loops for $\rho_{yx}^{odd}$ are extracted using Equation~\ref{Eq:spinningcurrents}: one for  $\mu_0H_{FC} = +1$ T case (Figure~\ref{summaryfig}e, yellow) and one $\mu_0H_{FC} = -1$ T case (Figure~\ref{summaryfig}f). In other words, the \gls{RMFR} extracts double the \textit{independent} data in the same time, or the same amount of \textit{independent} data in half the time. Comparing the data to that obtained via the antisymmetrization method in gray, again the only differences appear near the coercive fields, with origins the same as the \gls{ZFC} case.

The last scenario to discuss in Figure~\ref{flowchart} is Path 1-2-3-4-5'. Such a situation could be encountered in an exceptionally hard ferromagnet with an ordering temperature above that which can be achieved in the cryostat being used. In this hypothetical situation, the magnet used for applying the field would not have a large enough field to reach the coercive field to flip the ferromagnetic domains, and it would be impossible to heat high enough to thermally disorder the ferromagnet. Therefore, the \gls{TRS} necessary for using the antisymmetrization method could never be accessed. Here, so long as there are no processes that irreversibly damage the sample (like cracks caused by strain), the \gls{RMFR} method can still be used to measure the Hall effect.

With the two methods of extracting $\rho_{yx}^{odd}$ demonstrated on Co$_3$Sn$_2$S$_2$, we next use CeCoGe$_3$ as an example of how improper antisymmetrization can lead to artifacts that mimic the \gls{AHE}.

\section{C\MakeLowercase{e}C\MakeLowercase{o}G\MakeLowercase{e}$_3$}

\begin{figure}
\centering
\includegraphics[width = 0.8\textwidth]{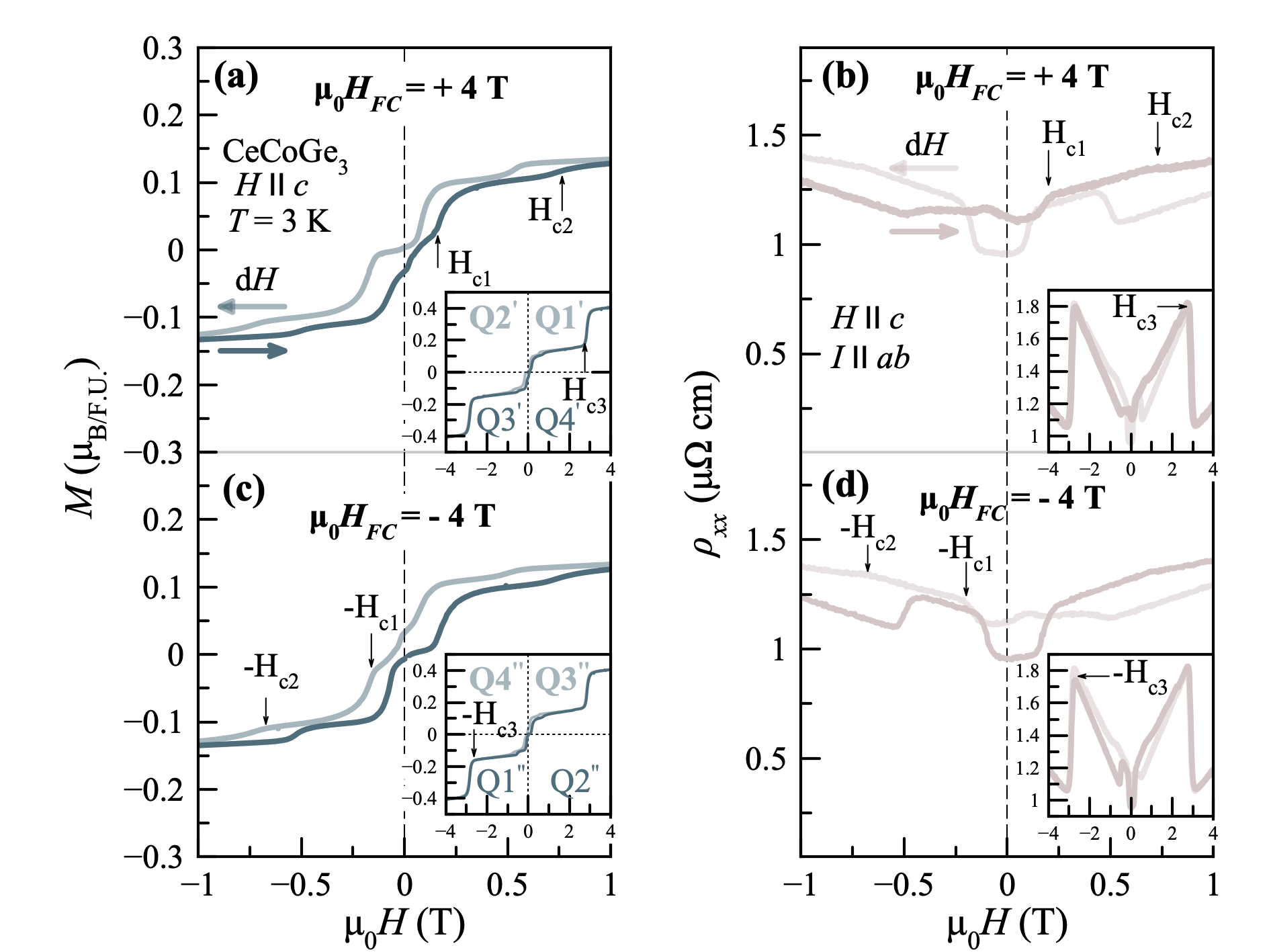}
\caption{\label{CeMRxxRyx} (\textbf{a}) Isothermal magnetization ($M$)  and (\textbf{b}) longitudinal resistivity ($\rho_{xx}$) measured at temperature $T$ = 3 K after cooling the sample from 100 K  in  $\mu_0H_{FC}~=~+4$ T using the positive field cool procedure. Q1'-Q4' in (\textbf{a}) label the sequential quarter-segments of the field sweep;  $+4~\text{T}\rightarrow 0$, $0\rightarrow -4~\text{T}$, $-4~\text{T}\rightarrow 0$, and $0\rightarrow +4~\text{T}$, respectively.  (\textbf{c}) $M$ and (\textbf{d}) $\rho_{xx}$ after field-cooling the sample in $\mu_0H_{FC}~=~-4$ T using the negative field cool procedure.  Q1$''$-Q4$''$ in (\textbf{c}) label the sequential quarter-segments of the field sweep;  $-4~\text{T}\rightarrow 0$, $0\rightarrow 4~\text{T}$, $4~\text{T}\rightarrow 0$, and $0\rightarrow -4~\text{T}$ (\textbf{c}) for the negative field-cool protocol. The main panels show data from $\mu_0H = \pm1$ T, while the insets show the same data from $\mu_0H = \pm4$ T. Light lines in all panels correspond to data measured from  $+4~\text{T}\rightarrow-4~\text{T}$ while dark lines correspond to $-4~\text{T}\rightarrow+4~\text{T}$. All measurements were performed with the applied magnetic field $H \parallel c$ while the transport measurements were performed with the current $I\parallel ab$.   Anomalies corresponding to metamagnetic transitions for increasing $H$ and $H>0$ or decreasing H and $H<0$ are labeled as $H_{c1}$, $H_{c2}$, $H_{c3}$ and $-H_{c1}$, $-H_{c2}$, $-H_{c3}$, respectively. Data used from Ref.~\cite{moya2026tunable}.} 
\end{figure}

\begin{figure}
\centering
\includegraphics[width = 0.7 \textwidth]{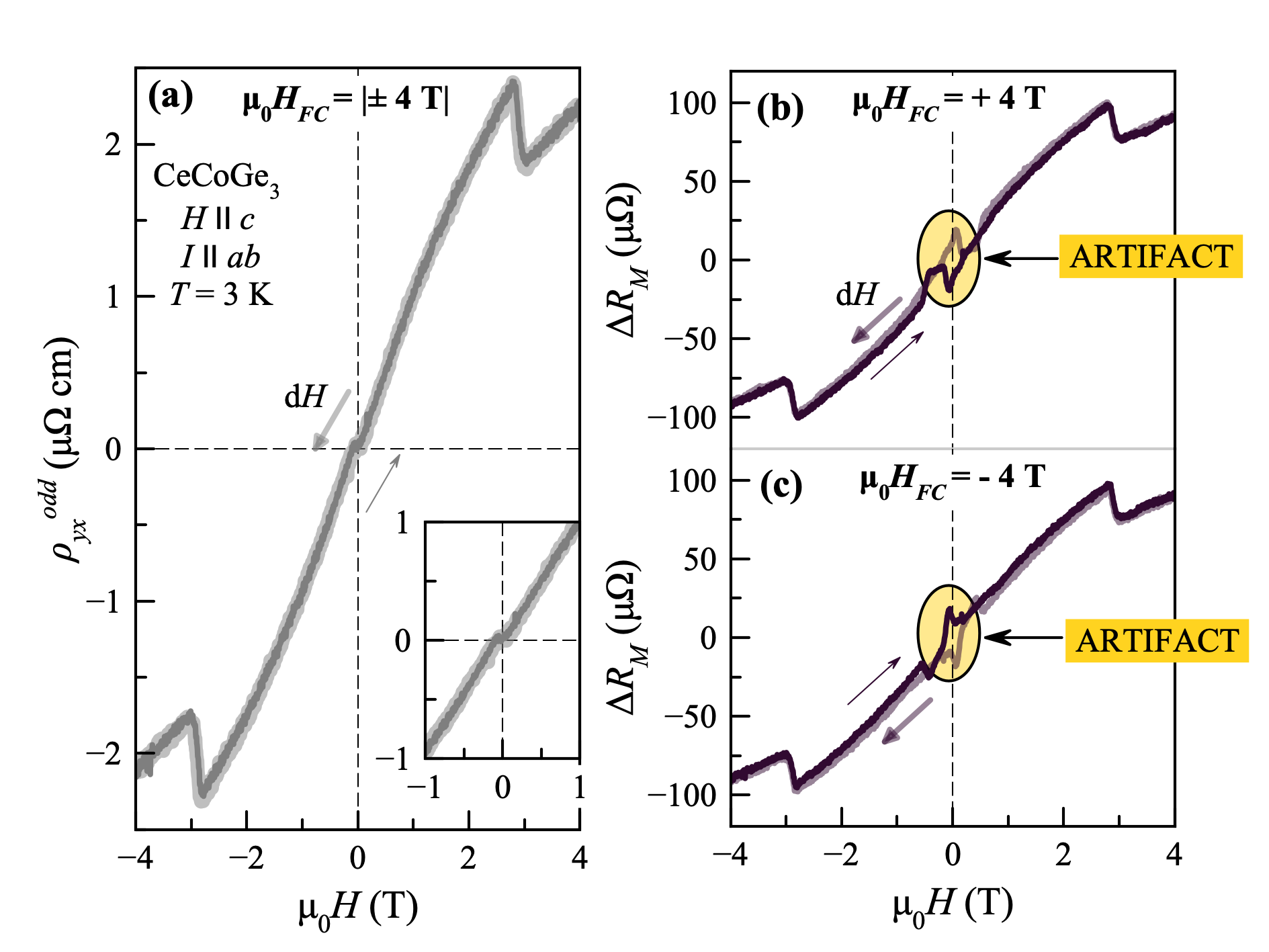}
\caption{\label{CeCOHall} (\textbf{a}) Isothermal Hall resistivity $\rho_{yx}^{odd}$ of CeCoGe$_3$ extracted using antisymmetrization with Eq. ~\ref{Eq:FC}  measured at temperature $T~=$ 3 K with magnetic field $H\parallel c$ and current $I \parallel ab$ after field cooling the sample in $\mu_0H_{FC}~=~|\pm 4|$ T. The inset is the same data zoomed in around zero-field showing there is no anomalous Hall effect. $\Delta R_m$ obtained by improper use of Equation~\ref{Eq:ZFC} under (\textbf{b}) the positive field-cool protocol with $\mu_0H_{FC}~=~+ 4$ T and (\textbf{c}) the negative field-cool protocol with $\mu_0H_{FC}~=~- 4$ T. Artifacts generated from improper antisymmetrization in (\textbf{b,c}) are highlighted in yellow.  Light lines in all panels correspond to data measured from  $+4~\text{T}\rightarrow-4~\text{T}$ while dark lines correspond to $-4~\text{T}\rightarrow+4~\text{T}$.}
\end{figure}

CeCoGe$_3$ crystallizes in the tetragonal,  non-centrosymmetric BaNiSn$_3$-type structure (space group $I4mm$) \cite{pecharsky1993unusual}, consistent with our X-ray diffraction experiments (Figure S1, Tables S5--S8). In this structure, the Ce atoms occupy the four corners and the body center of the tetragonal unit cell. The Co-Ge connectivity is asymmetric around the body-centered Ce along the crystallographic $c$-axis, breaking inversion symmetry \cite{pecharsky1993unusual}. This lack of centrosymmetry introduces the Dzyaloshinskii--Moriya (DM) interaction \cite{dzyaloshinsky1958thermodynamic,moriya1960new}, which competes with the Heisenberg-like RKKY interaction. Such competition can, in turn, lead to exotic domains or magnetic textures such as non-coplanar spin textures including skyrmions with application of a magnetic field \cite{fert2013skyrmions,tokura2010multiferroics,parkin2015memory,nagaosa2013topological}.

Indeed, the $H-T$ phase diagram of CeCoGe$_3$ with $H \parallel c$ has been found to be complex \cite{thamizhavel2005unique}. For $H = 0$, RKKY interactions induce long-range magnetic order below $T_N = 21$ K with a temperature-dependent evolution of magnetic structures \cite{pecharsky1993unusual, thamizhavel2005unique,smidman2013neutron}. The ground state, which exists below 8 K, is characterized by a magnetic propagation vector $\mathbf{k}$ (0,0,1/2) with a proposed two-up two-down magnetic structure \cite{smidman2013neutron}. With application of a magnetic field $H \parallel c$, three metamagnetic transitions have been reported below 8 K \cite{thamizhavel2005unique}. With competing energy scales and a complex $H-T$ phase diagram, CeCoGe$_3$ is a natural platform to search for non-coplanar spin textures, for which the Hall effect is a convenient first probe in most labs. Before presenting the Hall measurements, we first demonstrate that due to non-centered hysteresis caused by certain field-cooling procedures in $M$ and $\rho_{xx}$ about $H~=~0$, measuring the Hall effect in CeCoGe$_3$ is not straightforward.

Figure~\ref{CeMRxxRyx}a,b shows isothermal $M$ and $\rho_{xx}$, respectively, from Ref.~\cite{moya2026tunable}, measured for $H \parallel c$ using the positive field cool procedure  (Figure~\ref{summaryfig}b) with $T_{max}=100~\text{K}$, $T_{meas}=3~\text{K}$, and $\mu_0H_{max} = \mu_0H_{FC}~=~ 4$ T, where  $\mu_0H_{FC}$ is above the highest metamagnetic transition.  The low $H$ metamagnetic transitions for $Q4'$[Figure~\ref{CeMRxxRyx}a,b (main panel, dark lines)] are registered as anomalies in both $M$ and $\rho_{xx}$ near $\mu_0H_{c1}~\sim 0.16$ T and  $\mu_0H_{c2}~\sim 0.8$ T, while the higher $H$ metamagnetic transition near $\mu_0H_{c3}~\sim 3$ T is highlighted in the insets of Figures~\ref{CeMRxxRyx}a,b, consistent with previous measurements \cite{thamizhavel2005unique}. It is clear that the magnitude of $\mu_0H_{c1}$ and $\mu_0H_{c2}$ change between measurements taken $Q4'$ (dark lines) compared to the data taken with $Q1'$ (light lines) consistent with a first-order phase transition. More striking is the fact that $M(Q1') \neq -M(Q3')$ and $M(Q2') \neq -M(Q4')$  near $\mu_0H = 0$ T [Figure~\ref{CeMRxxRyx}a (main panel)]. Similarly, as seen in [Figure~\ref{CeMRxxRyx}b (main panel)],  a zero-field hysteresis is observed and $\rho_{xx}(Q1') \neq \rho_{xx}(Q3')$, $\rho_{xx}(Q2') \neq \rho_{xx}(Q4')$. The origins of this unusual asymmetry are due to glassy-like magnetic phase coexistence from a first-order phase transition and the details are reported elsewhere \cite{moya2026tunable}.

Nonetheless, while \gls{TRS} cannot be achieved in a single hysteresis loop, similar to the field-cooled cases of Co$_3$Sn$_2$S$_2$, following Path 1-2-3-4' in Figure~\ref{flowchart}, the \gls{TRS} can be found.  Figures~\ref{CeMRxxRyx}c,d show $M$ and $\rho_{xx}$, respectively, for the $\mu_0H_{FC}~=~ -4$ T (negative) field cool procedure. Comparing $M$ in Figure~\ref{CeMRxxRyx}a and \ref{CeMRxxRyx}c, it is apparent $M(Q1') = -M(Q1'')$, $M(Q2') = -M(Q2'')$, $M(Q3') = -M(Q3'')$, and $M(Q4') = -M(Q4'')$. Likewise $\rho_{xx}(Q1') = \rho_{xx}(Q1'')$, $\rho_{xx}(Q2') = \rho_{xx}(Q2'')$, $\rho_{xx}(Q3') = \rho_{xx}(Q3'')$, and $\rho_{xx}(Q4') = \rho_{xx}(Q4'')$, as can be seen in comparing Figure~\ref{CeMRxxRyx}b,d, confirming these respective states are indeed \gls{TRS} of each other. Therefore, if it is desired to extract $\rho_{yx}^{odd}$, using antisymmetrization, Equation~\ref{Eq:FC} (Path 1-2-3-4'-5 in Figure~\ref{flowchart})  should be applied to the raw data after taking the two measurements with the $\mu_0H_{FC}~=~+4$ T and  $\mu_0H_{FC}~=~-4$ T field-cooling procedures.

The corresponding Hall resistivity $\rho_{yx}^{odd}$ is presented in Figure~\ref{CeCOHall}a where the raw data was collected in a Hall bar geometry using the low-frequency \gls{AC} lock-in detection method and processed using the antisymmetrization method with Equation~\ref{Eq:FC}. $\rho_{yx}^{odd}$ displays a non-linear dependence on $H$, consistent with a multi-carrier response, shows little sensitivity to the metamagnetic transitions at $\mu_0H_{c1}$ and $\mu_0H_{c2}$, and exhibits a more pronounced change at $\mu_0H_{c3}$, consistent with recent reports \cite{furuhashi2025doping,shen2026giant}. Notably, $\rho_{yx}^{odd}(0)$ = 0 (no \gls{AHE}) even though $\rho_{xx}(0)$ and $M(0)$ show zero-field hysteresis. We confirmed that there is no \gls{AHE} in an independent crystal using \gls{RMFR} with a \gls{VdP} geometry shown in Figure S3. While the mobilities, and hence, the curvature of $\rho_{yx}^{odd}$ may vary from crystal to crystal, the lack of the \gls{AHE} and qualitative behavior $\rho_{yx}^{odd}$ are consistent. 

This scenario of no \gls{AHE} even though $\rho_{xx}(0)$ and $M(0)$ exhibit hysteresis is plausible if whatever mechanism is responsible for the \gls{AHE}, whether intrinsic or extrinsic averages to zero due to domain averaging in real-space, or is symmetry forbidden. We leave the exact mechanism to future studies, but move on to demonstrate, that incorrect treatment of the data can lead to a false \gls{AHE}.

If it was not realized that CeCoGe$_3$ exhibits non-centered hysteresis loops by measuring $M$ and $\rho_{xx}$ with the same field-cooling procedure, one may be tempted to use Equation~\ref{Eq:ZFC} which is only valid for centered hysteresis loops, to antisymmetrize the $R_m$ data. Doing so results in the traces shown in Figure~\ref{CeCOHall}b,c for $\mu_0H_{FC} = +4$ T and $\mu_0H_{FC} = -4$ T, respectively, where we have labeled the curves as $\Delta R_m$ to emphasize that $\Delta R_m$ does \textit{not} correspond to the true Hall effect $\rho_{yx}^{odd}$ in this case. If $\Delta R_m$ were incorrectly interpreted as $\rho_{yx}^{odd}$, a clear \gls{AHE} would be deduced in the zero-field limit, and it could be tempting to possibly attribute other $H>0$ features to the \gls{THE}. Given the hysteresis in $M$ and the non-centrosymmetric crystal structure, such a scenario could be possible. However, as already demonstrated in Figure~\ref{CeCOHall}a, correctly processing $R_m$ eliminates the possibility of an \gls{AHE} or \gls{THE} in CeCoGe$_3$. The artifact clearly originates from the hysteresis in $\rho_{xx}$. This comparison demonstrates the importance of correctly processing raw Hall data to eliminate artifacts and incorrect conclusions, and emphasizes the importance of measuring and reporting other thermodynamic or transport properties measured under the same $H-T$ history.

\section{Discussion}

\subsection{Artifacts-How to Spot Them and How to Correct Them}

As demonstrated, a central difficulty in extracting the Hall effect in  hysteretic materials is that measurements taken at opposite fields, $+H$ and $-H$, are not guaranteed to represent time-reversed states. In hysteretic materials $M(H)\neq -M(-H)$ for all $H$. In such cases a naive antisymmetrization at $\pm H$ in a single field sweep  can lead to artifacts like that shown in Figure~\ref{cartoon}a. Here, we have emphasized the idea of \gls{TRS}, so that Onsager--Casimir reciprocal relations which are strictly valid for states at $\pm B$ can be applied for the controlled variable $H$. Namely, Equation~\ref{Eq:paramagnetic} is the general antisymmetrization operation and remains valid so long as Onsager--Casimir reciprocity holds. We have discussed two methods - (1) \gls{RMFR} and (2) antisymmetrization with respect to applied field between \gls{TRS}, which throughout this manuscript we referred to simply as antisymmetrization - that can be used to access pairs of \gls{TRS}. Both methods are implementations of Equation~\ref{Eq:paramagnetic}; they differ only in how the \gls{TRS} are experimentally accessed.

\gls{RMFR} enforces time reversal by swapping current and voltage contacts between two otherwise identical measurements, which is equivalent to reversing the polarity of $B_z$ at the same applied field ($H_z$) setpoint. Therefore, \gls{RMFR}  provides a robust method of extracting the Hall effect in a single hysteresis loop whether or not the hysteresis is centered about $H = 0$. The only requirements for using the \gls{RMFR} method are that the sample and contacts are Ohmic and that the resistivity tensor obeys Onsager--Casimir reciprocity \cite{sample1987reverse}. For violations of these conditions we refer you to a recent review on non-linear transport \cite{suarez2025nonlinear}. Antisymmetrization, which has the same requirements, remains useful since it can be done with only one measurement channel, but care must be taken to assure the data is antisymmetrized between only \gls{TRS}.  We have provided a decision tree and field cooling protocols to explain how \gls{TRS} states can be achieved in Figure~\ref{summaryfig}. 

The comparison of the two methods on Co$_3$Sn$_2$S$_2$ clarifies the trade-offs between the two methods. In the \gls{ZFC} case where the hysteresis loops are centered, we have followed Path 1-2-3-4 in Figure~\ref{flowchart}. Piecewise antisymmetrization reproduces the \gls{RMFR} result away from coercivity, confirming $\rho_{yx}^{odd}$ is captured correctly when time-reversal pairing is respected (Figure~\ref{Co3HallM}e). Once \gls{EB} is introduced by field cooling (Path 1-2-3-4'-5 in Figure~\ref{flowchart}), the loop is no longer centered about $H=0$ (Figure~\ref{summaryfig}e,f). \gls{RMFR} continues to isolate $\rho_{yx}^{odd}$ reliably in a single hysteresis loop, while antisymmetrization requires over double the time since two hysteresis loops of $R_m$ need to be measured in reverse field-cool procedures with respect to each other where the sample would have to be warmed above the ordering temperature and the field oscillated to zero between. These facts highlight two advantages of the \gls{RMFR} method -  $\rho_{yx}^{odd}$ can be captured over twice as fast and no a priori knowledge of whether the hysteresis loops are centered or not is needed.

The two methods diverge, most notably as kinks near the coercive fields. Here, one typically finds a mundane cause: residual field in the superconducting magnet subtly coupled with a sharp feature in magnetoresistance, imprinting features that antisymmetrization cannot cancel (Figure~\ref{summaryfig}d, inset). The \gls{RMFR} method inherently mitigates the problem because it measures an $R$ equivalent to exactly an opposite $B$ at the same nominal $H$. More broadly, anomalies around the coercive fields warrant caution: features such as bumps are often attributed to a topological Hall contribution~\cite{yasuda2016geometric,liu2017dimensional,ludbrook2017nucleation,matsuno2016interface}, while alternative explanations for some cases have been proposed~\cite{gerber2018interpretation,tai2022distinguishing}.  Regardless of the underlying physics, minimizing instrumental artifacts is essential for interpretation.

While we focused on magnetic materials in this work, it should be noted that trapped flux together with a large magnetoresistance, can cause artifacts in Hall measurements when using antisymmetrization methods even in non-magnetic materials. If enigmatic Hall data is measured in such a scenario, it is worth checking a measurement using \gls{RMFR} to make sure the data is correct.

The CeCoGe$_3$ case illustrates the broader lesson for materials with glassy or coexisting magnetic domains with strong pinning. Such a scenario may happen when there is magnetic frustration, disorder, magnetic phase coexistence, or near first-order phase transitions. Here, we demonstrated the importance of reporting $M(H)$ and the longitudinal resistivity $\rho_{xx}(H)$ (Figure~\ref{CeMRxxRyx}) together with $\rho_{yx}^{odd}$ (Figure~\ref{CeCOHall}a). In measuring $\rho_{xx}(H)$ and $M(H)$ we were able to deduce the apparent  \gls{AHE}-like signature registered in $\Delta R_m$ (Figure~\ref{CeCOHall}b,c) arises from asymmetric $\rho_{xx}(H)$.  

\subsection{Contact Geometry}\label{Contact}

The methods presented here only work for extracting the Hall effect, $\rho_{yx}^{odd}$. In materials with low enough symmetry as described in Section~\ref{details}, $\rho_{yx}$ can have even-in-$B_z$ anisotropic magnetoresistive terms which are necessarily canceled by both the \gls{RMFR} and antisymmetrization techniques. Careful consideration of the contact geometry would be needed to extract the full resistivity tensor.

When only the Hall effect, $\rho_{yx}^{odd}$, is desired, the Onsager--Casimir reciprocal relation \cite{onsager1931reciprocal,onsager1931reciprocal2,casimir1945onsager,akgoz1975space}  $\rho_{yx}^{odd}(B_z) = -\rho_{xy}^{odd}(B_z)$, implies $\rho_{yx}^{odd}$ is independent of current direction, and therefore can be extracted with arbitrary contact placement, as long as $B_z$ is parallel to the face-normal of the $xy$-plane, and the voltage contacts are placed in such a way to pick up some transverse signal. 

The choice of whether to choose between, for example, a \gls{VdP} geometry or Hall bar geometry depends on how large the transverse voltage is relative to the longitudinal voltage. For reference of what these geometries can look like in practice, images of the contact geometries used for measurements on Co$_3$Sn$_2$S$_2$ are shown in Figure S4.  If the transverse voltage is large, either geometry is equally valid, but a \gls{VdP} geometry has the advantage of being able to measure the anisotropic resistivity via the Montgomery method \cite{montgomery1971method,dos2011procedure} and the Hall effect with the same contacts. On the other hand, if the transverse voltage is small compared to the longitudinal voltage, a Hall bar geometry can be used to minimize the relative size of the longitudinal signal in the measured signal. This comes at the cost of having to put more contacts on to measure any resistivity anisotropies.

Also, $\rho_{yx}^{odd}$ can be obtained by multiplying $R_{yx}^{odd}$ by the thickness of the crystal in either \gls{VdP} or Hall bar geometry. The assumptions are that the sample is  uniform and ohmic with well-defined thickness and current path. 

\subsection{Inhomogeneities}

Importantly, the assumption of uniform thickness and homogeneity is essential to convert from $R_{yx}^{odd}$ to $\rho_{yx}^{odd}$.  Historically, Hall measurements have also been used as a practical check of sample homogeneity since different contact configurations probe different current paths through the specimen. If a sample is homogeneous and the geometry is well controlled, those configurations (assuming they are in the same measurement plane) should be mutually consistent up to the expected geometric factors as discussed in Section~\ref{Contact}. If they are not, the discrepancies can themselves reflect inhomogeneity \cite{sample1987reverse}. Namely, if the spread between configurations is larger than the propagated measurement uncertainty and expected geometric-factor variation, this is consistent with sample inhomogeneity.

It is then important to distinguish between what \gls{RMFR} does and does not guarantee. \gls{RMFR} states that, for a given four-contact measurement, the ordered reciprocal configuration measured at reversed magnetic induction is equal i.e. 
$R_{12,34}(+B_z)=R_{34,12}(-B_z).$ \cite{sample1987reverse,cornils2008reverse}. It does not imply that two different choices of contacts on an inhomogenous sample must yield the same $R_{yx}^{odd}$ value. Rather, each contact configuration obeys \gls{RMFR} with its own reciprocal partner.

In this sense, both antisymmetrization and \gls{RMFR} isolate the odd-in-field Hall response for a chosen measurement geometry. They remove the usual even-in-field contamination, such as longitudinal pickup from contact misalignment.  If different contact choices yield different Hall responses, that may reflect genuine sample inhomogeneity and  $R_{yx}^{odd}$ should not be converted to $\rho_{yx}^{odd}$.

\subsection{Anomalous and Topological Hall effects}
We only covered how to reliably \textit{measure} the total Hall effect $\rho_{yx}^{odd}$. Historically, $\rho_{yx}^{odd}$ = $\rho_{yx}^{N}+\rho_{yx}^{AHE}+\Delta\rho_{yx}$, has been divided into three contributions. The first term $\rho_{yx}^{N}$ is the normal Hall effect \cite{hall1879new} which arises due to the Lorentz force electrons experience in a magnetic field. $\rho_{yx}^{AHE}$ arises from reciprocal-space Berry curvature, skew scattering or side-jump scattering, and typically depends on $M$. $\Delta\rho_{yx}$ are terms that do not depend on $M$. For detailed discussions, analysis and interpretation of $\rho_{yx}^{AHE}$, we refer to Ref.~\cite{nagaosa2010anomalous,xiao2010berry}. $\Delta\rho_{yx}$ can either come from the real-space topology of magnetic structures, sometimes called the \gls{THE} or more generally from net \gls{SSC} in non-coplanar spin textures, for which real-space topological spin textures are a special case. Here, we refer to Refs. \cite{bruno2004topological,taguchi2001spin,tokura2020magnetic,wang2022topological}. Finally, Hall effects can emerge in antiferromagnets with exceedingly small magnetization that are not related to $\rho_{yx}^{N}$, and therefore, by the previous definition, would fall into the $\Delta\rho_{yx}$ category. However, this phenomenon is referred to as the \gls{AHE} since its origins are reciprocal-space Berry curvature as reviewed in \cite{vsmejkal2022anomalous}. The physics and history of measuring the Hall effect is vast and quickly evolving. We therefore do not attempt a complete list of relevant references. However, it is worth cautioning not to overinterpret $\Delta\rho_{yx}$ since there can be many origins \cite{gerber2018interpretation,kimbell2022challenges,kurumaji2024metamagnetic,rai2025cusplike}.

\subsection{Relevance to Transition-Metal Intercalated Transition Metal Dichalcogenides}

The Hall analysis procedures described here are generic and applicable to any conductor, but they are indispensable in hysteretic materials. As noted in the Introduction, \gls{EB} was first associated with engineered AFM/FM thin-film heterostructures \cite{nogues1999exchange}, but it has recently  been shown in single-crystalline magnetic conductors as well \cite{lachman2020exchange,noah2022tunable,xu2022ferromagnetic,firdosh2025exchange,kotegawa2023large,maniv2021exchange}. A particularly promising platform for exploring \gls{EB}, or more generally, non-centered hysteresis and related functionalities is the family of transition-metal--intercalated \gls{TMDs}. Parkin and Friend showed in 1980 that magnetic intercalation in \gls{TMDs} produces highly anisotropic layered magnets with large \gls{AHE} responses \cite{parkin19803,parkin19804,parkin1980magnetic}. Subsequent work has revealed a rich variety of spin textures and domain states controlled by host material and intercalant concentration. Representative examples include \gls{AHE} or \gls{THE} signals from non-coplanar spin textures in Cr$_{1/3}$NbS$_2$, Co$_{1/3}$TaS$_2$, and Cr$_{1/3}$TaS$_2$ \cite{miyadai1983magnetic,togawa2012chiral,clements2017critical,zhang2021chiral,goodge2023consequences,park2023tetrahedral,takagi2023spontaneous}, as well as sharp magnetic switching accompanied by an \gls{AHE} in the hard ferromagnet Fe$_{1/4}$TaS$_2$ \cite{morosan2007sharp,checkelsky2008anomalous,chen2016correlations,hardy2015very}. These materials therefore provide a natural arena where Hall measurements directly probe magnetic order and domain physics.

At the same time, their sensitivity to stoichiometry and disorder makes the phase space highly tunable, but prone to non-centered hysteresis. Phase coexistence and disorder are precisely the ingredients that can generate non-centered hysteresis loops, especially if measurements are done in the field-cooled state. For example, Fe$_{1/3+\delta}$NbS$_2$ develops a coupled spin-glass/antiferromagnetic state that yields giant \gls{EB} and enables energy-efficient electrical switching \cite{nair2020electrical,maniv2021exchange}, with magnetic domains acutely sensitive to stoichiometry \cite{wu2022highly}. V$_{1/3}$NbS$_2$ hosts a multidomain state producing a spontaneous \gls{AHE} emerging from a non-Fermi-liquid regime \cite{ray2025zero}, while Cr$_{1/4}$TaS$_2$ shows \gls{AHE} and glassy behavior associated with magnetic phase coexistence \cite{xie2025anomalous}. Room-temperature exchange bias has also been reported in Fe$_{0.17}$ZrSe$_2$ \cite{kong2023near}. Given that intercalated \gls{TMDs} already exhibit both exotic spin textures and \gls{EB}, similar physics is likely across other compositions. The methods presented enable measurement of the Hall effect in the \gls{EB} state, and will be exceptionally useful to distinguish intrinsic effects from measurement artifacts in this material family as well as others with similar interplay of complex magnetism, stoichiometry, and disorder.

\section{Conclusions and Outlook}\label{sec5}

To conclude, we have shown how extracting the true odd-in-field Hall response in hysteretic magnetic materials is not as simple as antisymmetrizing $R_m(+H)$ with $R_m(-H)$ because opposite applied fields do not necessarily prepare time-reversed magnetic states once there is coercivity, exchange bias, glassiness, or multi-step metamagnetism. As a result, incorrect processing can generate \gls{AHE}- or \gls{THE}-like artifacts that result from uncompensated longitudinal signals in the measured voltages. We addressed this by laying out two practical routes: (1) the \gls{RMFR} method and (2) antisymmetrization of time-reversed states under applied magnetic field. We provide a workflow to record intrinsic Hall voltages from measurements. These methods are general and will continue to be useful in the characterization of magnetic conductors well beyond the examples discussed here.

Broadly, Hall measurements have already provided enormous insight into magnetic and topological materials \cite{nagaosa2010anomalous,tokura2020magnetic,vsmejkal2018topological}. Looking forward they will remain an essential tool as research activity continues to grow in areas such as altermagnetism and  multipolar higher-order magnetic states \cite{vsmejkal2022anomalous,gonzalez2023spontaneous,yamada2025metallic}. In these systems, the presence, absence, or detailed form of the \gls{AHE} can itself be highly informative, and in some cases a Hall response in a compensated or very small-moment system is precisely what makes the material attractive for spintronic functionality \cite{vsmejkal2022anomalous,yamada2025metallic}. For these reasons, practical methods that distinguish intrinsic Hall signals from measurement artifacts should become increasingly important as the search for unconventional magnetic responses and useful spintronic materials continues.

\section{Methods}

\subsection{Crystal Growth}

Co$_3$Sn$_2$S$_2$ was grown out of a ternary melt \cite{lin2012development}. The constituent elements were weighed in the ratio Co$_{12}$Sn$_{80}$S$_{6}$, put in an alumina crucible, and sealed in an evacuated quartz tube. The ampoule was heated to $400^{\circ}\mathrm{C}$ over 2~h and held for 2~h, then ramped to $1050^{\circ}\mathrm{C}$ over 6~h and dwelled for 10~h. It was subsequently cooled to $700^{\circ}\mathrm{C}$ over 60~hrs, at which point the remaining flux was separated by centrifugation.

CeCoGe$_3$ was grown from Bi flux \cite{austingrowth,thamizhavel2005unique}. Ce:Co:Ge were weighed in the stoichiometric molecular ratio of 1:1:3 and arc melted in an Ar atmosphere. The mass loss after arc melting was 2.8$\%$. The arc melted bead was then put in an alumina crucible with Bi such that the total loaded composition was Ce$_6$Co$_6$Ge$_{18}$Bi$_{70}$.  The growth was sealed in an evacuated quartz ampoule, heated to 1150$^\circ \text{C}$ at 120$^\circ \text{C}/$h where it dwelled for 5 h, before cooling to 650$^\circ \text{C}$ over 148 h where it was centrifuged to separate the crystals from the flux.

\subsection{X-Ray Diffraction}

Single crystal X-ray diffraction measurements were performed at room temperature or 100 K using a Rigaku XtaLAB Synergy-S/i diffractometer equipped with a Mo K$_{\alpha1}$ ($\lambda$ = 0.71073 \AA), microfocus sealed-tube X-ray source and a graphite monochromator. Integration was carried out using CrysAlis$^{\text{Pro}}$ software, with numerical absorption correction based on Gaussian integration over a multifaceted crystal model. Full structural refinements on F$^2$ were performed in JANA2020.

\subsection{Electrical Transport}

A single crystal of Co$_3$Sn$_2$S$_2$ was shaped into an approximate rectangular plate. For the Hall measurements  on Co$_3$Sn$_2$S$_2$  electrical contacts were attached in the $ab-$ plane in a \gls{VdP} geometry using silver paint and gold wires. Measurements were done in a Quantum Design (QD) DynaCool physical properties measurement system (PPMS) cryostat using a Lake Shore M81-SSM synchronous source measure system equipped with BCS-10 balanced current source modules and VM-10 voltage measure modules. The two current sources were set with 10 mA amplitudes running at different frequencies (17.17 and 7.717 Hz). The two voltage measurement modules were locked into the respective frequencies while setting the ground to the voltage sink in m1 (V-), or equivalently, the current sink in m2 (I-) (Figure~\ref{summaryfig}c) using the Common-Mode Rejection feature of the M81-SSM allowing for simultaneous acquisition of m1 and m2. The magnetic field was swept at 50 Oe/s. The independent Hall bar measurement in Figure~\ref{summaryfig}d was measured on the same crystal. The methods for extracting the Hall resistivity are discussed in the paper.  Because the Hall voltage is so large in Co$_3$Sn$_2$S$_2$, a collinear geometry was used for measurements of the longitudinal resistivity. Even still, the raw signal must be symmetrized using

\begin{equation}
R_{xx}(H) = \begin{cases}
\frac{+(+)R_m(Q1)+(+)R_m(Q3)}{2} & \text{for}~H[Q1,(Q3)] \\
\frac{+(+)R_m(Q2)+(+)R_m(Q4)}{2} & \text{for}~H[Q2,(Q4)], \\
\label{Eq:ZFC_rxx}
\end{cases}
\end{equation}

\noindent for the \gls{ZFC} case, and 

\begin{equation}
R_{xx}(H) = \begin{cases}
\frac{+(+)R_m(Q1')+(+)R_m(Q1'')}{2} & \text{for}~H[Q1',(Q1'')] \\
\frac{+(+)R_m(Q2')+(+)R_m(Q2'')}{2} & \text{for}~H[Q2',(Q2'')] \\
\frac{+(+)R_m(Q3')+(+)R_m(Q3'')}{2} & \text{for}~H[Q3',(Q3'')] \\
\frac{+(+)R_m(Q4')+(+)R_m(Q4'')}{2} & \text{for}~H[Q4',(Q4'')], \\
\label{Eq:FC_rxx}
\end{cases}
\end{equation}

\noindent for the \gls{EB} cases.

Since the Hall voltage is small compared to the longitudinal voltage in CeCoGe$_3$ we opted for a Hall bar geometry where electrical contacts were attached in the $ab-$ plane using gold paint and the signals were acquired using the low-frequency lock-in detection method ($f$ = 15.259 Hz with a $I$ = 5 mA current) in a QD DynaCool PPMS  using the ETO option. The magnetic field was swept at 25 Oe/s. Hall data was antisymmetrized using the methods described in the text. Magnetoresistance was measured with the same parameters  in a collinear geometry and was not symmetrized.

\subsection{Magnetization}

Magnetization measurements were performed in a QD magnetic measurement system (MPMS) equipped with the VSM-SQUID option. Samples were mounted on either a brass holder for Co$_3$Sn$_2$S$_2$, or a quartz holder for CeCoGe$_3$ using GE varnish. The magnetic field was swept at 50 Oe/s or 25 Oe/s respectively.

% \backmatter
\section*{Author Contributions}

J.M.M. and L.M.S. conceived of the project idea. J.M.M. and A.V. grew the crystals and performed the magnetic and electrical transport measurements and analysis. J.M.M. performed single-crystal X-ray diffraction experiments and S.B.L. solved the crystal structures. S.C. provided significant insights on how to measure m1 and m2 simultaneously. G.S. made J.M.M. aware of the potential usefulness of the \gls{RMFR} method. C.J.P. provided valuable insights about the symmetry conditions in the off-diagonal components of the resistivity tensor. L.M.S. supervised the work. J.M.M. and L.M.S. wrote the manuscript with input from all authors.

\section*{Acknowledgments}
This work was supported by the Air Force Office of Scientific Research under award numbers A9550-23-1-0635 and FA9550-25-1-0177. Additional support was provided by an NSF CAREER grant (DMR-2144295) to LMS and the Gordon and Betty Moore Foundation's EPIQS initiative through Grant GBMF9064. We thank Jeffrey Lindemuth and Emilio A. Codecido for useful discussions.

\section*{Conflicts of Interest}

The authors declare no conflicts of interests.

\section*{Data Availability Statement}

Thermodynamic and transport data first reported in this manuscript are publicly available on Zenodo at \url{https://doi.org/10.5281/zenodo.20312615}. Data that appear in related manuscripts are cited where used and are not included in this public repository.

% Reset the bibliography heading one more time immediately before BibTeX output.
\renewcommand{\bibsection}{\section*{References}}
\bibliographystyle{unsrtnat}
\bibliography{wileyNJD-APA_formanuscript}

\clearpage
\appendix
%%%%%%%%%%%%%%%% START OF SUPPLEMENT %%%%%%%%%%%%%%%

% Figures, tables, equations and pages in the supplement are numbered S1, S2 etc.
\renewcommand{\thefigure}{S\arabic{figure}}
\renewcommand{\thetable}{S\arabic{table}}
\renewcommand{\theequation}{S\arabic{equation}}
% Avoid duplicate hyperref anchors after resetting counters in the supplement.
\renewcommand{\theHfigure}{S\arabic{figure}}
\renewcommand{\theHtable}{S\arabic{table}}
\renewcommand{\theHequation}{S\arabic{equation}}
\setcounter{figure}{0}
\setcounter{table}{0}
\setcounter{equation}{0}
% \setcounter{page}{1} % page numbering continues in this reformatted version
% References continue the numbering from the main text.

%%%%%%%%%%%%%%%% SUPPLEMENT TITLE PAGE %%%%%%%%%%%%%%%

\newpage
% one-column layout already active in article.cls
\section*{Supplementary Materials for: Measuring the Hall effect in hysteretic materials}
\begin{center}
% Author list for the supplement
% Indicate the corresponding authors, but do NOT include institutions here
% It would be nice if the template auto-generated this, but doing so is complicated...
Jaime M. Moya,
    Anthony Voyemant,
    Sudipta Chatterjee,
    Scott B. Lee,
    Grigorii Skorupskii,
    Connor J. Pollak,
    and Leslie M. Schoop

\end{center}

\begin{figure}[htbp]
\centering
\includegraphics[width = \textwidth]{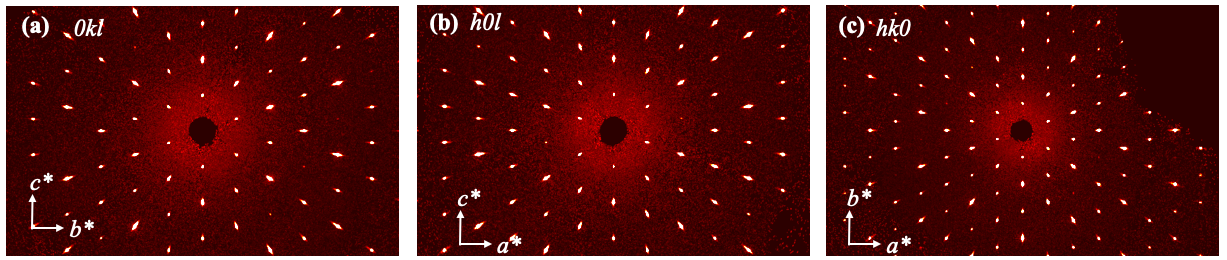}
\caption{\label{Co3Sn2S2Precession} Precession images of Co$_3$Sn$_2$S$_2$ in the  (\textbf{a}) $0kl$, (\textbf{b}) $h0l$ and (\textbf{c}) $hk0$ planes measured at 300 K. \label{Co3Sn2S2xrd} }
\end{figure}

\begin{figure}[htbp]
\centering
\includegraphics[width = \textwidth]{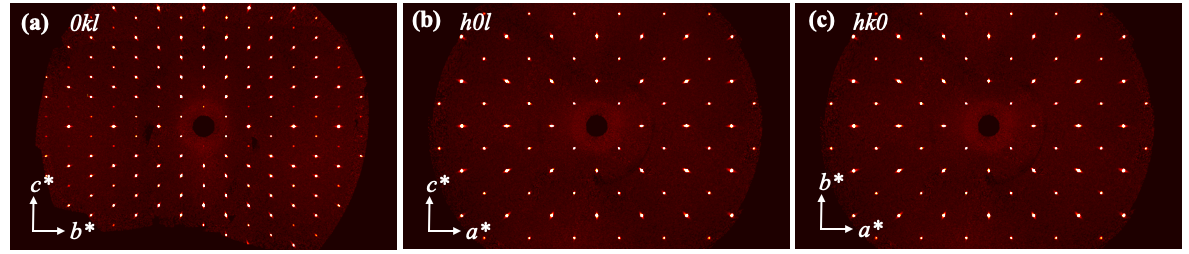}
\caption{\label{CeCoGe3_xrd} Precession images of CeCoGe$_3$ in the  (\textbf{a}) $0kl$, (\textbf{b}) $h0l$ and (\textbf{c}) $hk0$ planes measured  at 100 K.  }
\end{figure}

%%% Crystallographic Refinement Information 
\begin{table} [hbt!]
\renewcommand{\tablename}{Table S}
\caption{Crystallographic Information for Co$_3$Sn$_2$S$_2$.}
 \centering
 \begin{tabular}{l c}
\hline
Refined Composition & Co$_3$Sn$_2$S$_2$ \\
\hline 
Crystal Dimension (mm) &  0.084 $\times$ 0.072 $\times$ 0.024  \\
Radiation Source, $\lambda$ (\AA) & { Mo K$_{\alpha}$, 0.71073 }   \\
Absorption Correction &  gaussian   \\
Data Collection Temperature (K)   &  297.84(10) \\
Space Group &  R$\overline{3}m$    \\
$a$ (\AA) &  5.36979(10)   \\
$c$ (\AA) &  13.1835(3)  \\
Cell Volume (\AA$^{3}$)  & 329.212(11) \\
Absorption Coefficient (mm$^{-1}$) & 23.068 \\
$\theta_{min}$ , $\theta_{max}$ (deg) &  4.64, 44.87 \\
Refinement Method & F$^{2}$  \\
R$_{int}$(I\textgreater{}3$\sigma$, all)   &  8.43, 8.44    \\
Number of Parameters     &  13    \\
Unique Reflections (I \textgreater 3$\sigma$, all)  & 345, 411  \\
R(I\textgreater{}$3\sigma$), R$_{w}$(I\textgreater{}$3\sigma$)   &   1.48, 4.09    \\
R(all), R$_{w}$(all)      & 1.54, 4.13   \\
S(I\textgreater{}$3\sigma$), S(all)   & 1.2910, 1.2667   \\
$\Delta\rho_{max}$ , $\Delta\rho_{min}$ (e \AA$^{-3}$) & 0.45, -0.70  \\ 
\hline
\label{CI_CSS}
\end{tabular}
\end{table}

%%%
%%% Crystallographic Atomic Coordinates
%%%
\begin{table}[hbt!]
\renewcommand{\tablename}{Table S}
\caption{Refined atomic coordinates for Co$_3$Sn$_2$S$_2$.}
\begin{tabular}{@{}lcccccc@{}}
\hline
Site & Wyckoff Position & x   & y    & z          & Occupancy &  \\  
\hline
Sn1 & 3a &  2/3 &  1/3 & 1/3  & 1 & \\
Sn2 & 3b &  1/3 &  2/3 &  1/6  & 1 &  \\
Co1 & 9e &  1/6 &  1/3 &  1/3  & 1  & \\
S1  & 6c  & 1/3  & 2/3  & 0.44952(5) & 1 & \\ 
\hline
\label{AC_CSS}
\end{tabular}
\end{table}

%%%
%%%  Anisotropic Parameters
%%%
\begin{table}[hbt!]
\renewcommand{\tablename}{Table S}
\caption{Refined anisotropic displacement parameters for Co$_3$Sn$_2$S$_2$.}
\begin{tabular}{lccccccc}
\hline
Site & U$_{11}$ & U$_{22}$   & U$_{33}$    & U$_{12}$ & U$_{13}$         & U$_{23}$ &  \\
\hline
 Sn1 & 0.00577(8)  & 0.00577(8)  & 0.01288(12) & 0.00289(4) & 0           &  0 & \\
 Sn2 & 0.00972(9)  & 0.00972(9)  & 0.00502(10) & 0.00486(4) & 0           &  0 & \\
 Co1 & 0.00745(11) & 0.00580(13) & 0.00752(13) & 0.00290(6) & -0.00054(4) & -0.00108(9) & \\
 S1  & 0.00711(17) & 0.00711(17) & 0.0054(2)   & 0.00356(8) & 0           &  0 & \\
\hline 
\label{AP_CSS}
\end{tabular}
\end{table}
%%%
%%% Distances
%%%
\begin{table}[hbt!]
\renewcommand{\tablename}{Table S}
\caption{Selected interatomic distances for Co$_3$Sn$_2$S$_2$.}
\begin{tabular}{llclclclcl}
\hline
 Site && Neighbor && Multiplicity   && Distance (\AA) &&  \\
\hline
  Sn1 && Co1 && 6 &&  2.68490(10) && \\
  Sn2 &&  Co1 && 6 &&  2.68901(9) &&  \\
  Co1 && S1 && 2 &&  2.1793(5) &&  \\
\hline
\label{D_CSS}
\end{tabular}
\end{table}

\begin{table} [ht!]
\renewcommand{\tablename}{Table S}
\caption{\textbf{Crystallographic Data} for the refinement of CeCoGe$_3$.}
 \centering
 \begin{tabular}{l c c}
\hline
Refined Composition & CeCoGe$_3$ & CeCoGe$_3$ \\ 
\hline 
Crystal Dimension (mm) & \multicolumn{2}{c}{0.062 $\times$ 0.073 $\times$ 0.090} \\
Radiation source, $\lambda$ (\AA) & \multicolumn{2}{c}{X-ray, 0.7109} \\
Absorption Correction & \multicolumn{2}{c}{multi-scan} \\
Data Collection Temperature (K)  & 100 & 295 \\
(3+1) D superspace Group & \multicolumn{2}{c}{I4\textit{mm}} \\
$a$ (\AA) & 4.30949(16) & 4.32185(11) \\
$b$ (\AA) & 4.30949(16) & 4.32185(11) \\
$c$ (\AA) & 9.8186(3) &  9.8396(2) \\
Cell Volume (\AA$^{3}$)  &  182.348(11) &  183.788(8) \\
Absorption Coefficient (mm$^{-1}$) & \multicolumn{2}{c}{40.856} \\
$\theta_{min}$ , $\theta_{max}$ (deg) & 4.15, 37.73 & 4.14, 37.9\\
Refinement Method & \multicolumn{2}{c}{F$^{2}$}   \\
R$_{int}$(I\textgreater{}3$\sigma$, all) & 12.49, 12.49 & 11.42, 11.42 \\
Total Reflections (I \textgreater 3$\sigma$, all)  & 0, 6071 & 6229, 6376 \\
Unique Reflections (I \textgreater 3$\sigma$, all)  & 320, 320 & 326, 326  \\
Number of Parameters    & 10 & 15 \\
R(I\textgreater{}$3\sigma$), R$_{w}$(I\textgreater{}$3\sigma$)  &  4.55,  11.27 & 6.06, 13.50 \\
R(all), R$_{w}$(all)     &   4.55, 11.27  & 6.06, 13.50 \\
S(I\textgreater{}$3\sigma$), S(all)   & 4.7497, 4.7497 & 5.2155, 5.2155 \\
$\Delta\rho_{max}$ , $\Delta\rho_{min}$ (e \AA$^{-3}$) & 1.51, -1.09 & 2.50, -1.07 \\
\hline 
\label{CI_xtalA}
\end{tabular}
\end{table}

\begin{table} [ht!]
\renewcommand{\tablename}{Table S}
\caption{\textbf{Atomic positions} for the refinement of CeCoGe$_3$ at 100K.}
 \centering
 \begin{tabular}{c c c c c c}
\hline
Label  & Element  & $x$ &  $y$  & $z$ &  U$_{eq}$ \\
\hline
  Ce1 & Ce & 0 & 0  & 0.43392(8) & 0.0006(3) \\
  Ge1 & Ge & 0.5  & 0.5 & 0.5014(2) & 0.0012(3) \\
  Ge2 & Ge & 0.5  & 0 & 0.17388(15) & 0.0015(3) \\
  Co1 & Co & 0.5  & 0.5 & 0.2680(3) & 0.0011(5) \\
\hline
\label{Positions_A_100}
\end{tabular}
\end{table}

\begin{table} [ht!]
\renewcommand{\tablename}{Table S}
\caption{\textbf{Atomic positions} for the refinement of CeCoGe$_3$ at RT.}
 \centering
 \begin{tabular}{c c c c c c}
\hline
Label  & Element  & $x$ &  $y$  & $z$ &  U$_{eq}$ \\
\hline
Ce1 & Ce & 0   &  0   &   0.53886(11) & 0.0072(3) \\
Ge1 & Ge & 0.5 &  0.5 &   0.4708(2)   & 0.0072(5) \\
Ge2 & Ge & 0.5 &  0   &   0.2988(2)   & 0.0083(5) \\
Co1 & Co & 0.5 &  0.5 &   0.7042(4)   & 0.0066(7) \\
\hline
\label{Positions_A_RT}
\end{tabular}
\end{table}

\begin{table} [ht!]
\renewcommand{\tablename}{Table S}
\caption{\textbf{Components of the anisotropic ADP (U$_{ani}$) parameters} for the refinement of Crystal CeCoGe$_3$ 295 K.}
 \centering
 \begin{tabular}{c c c c c c c}
\hline
Atom & U$_{11}$ & U$_{22}$ & U$_{33}$ & U$_{12}$ & U$_{13}$ & U$_{23}$ \\
\hline
 Ce1 & 0.0071(4) & 0.0071(4)   &  0.0073(6)  & 0 & 0 & 0 \\
 Ge1 & 0.0082(5) & 0.0082(5)   &  0.0051(14) & 0 & 0 & 0 \\
 Ge2 & 0.0062(9) & 0.0121(10)  &  0.0067(9)  & 0 & 0 & 0 \\
 Co1 & 0.0061(8) & 0.0061(8)   &  0.0074(16) & 0 & 0 & 0 \\
\hline
\label{U_A_RT}
\end{tabular}
\end{table}

\begin{figure}
\centering
\includegraphics[width = 0.5\textwidth]{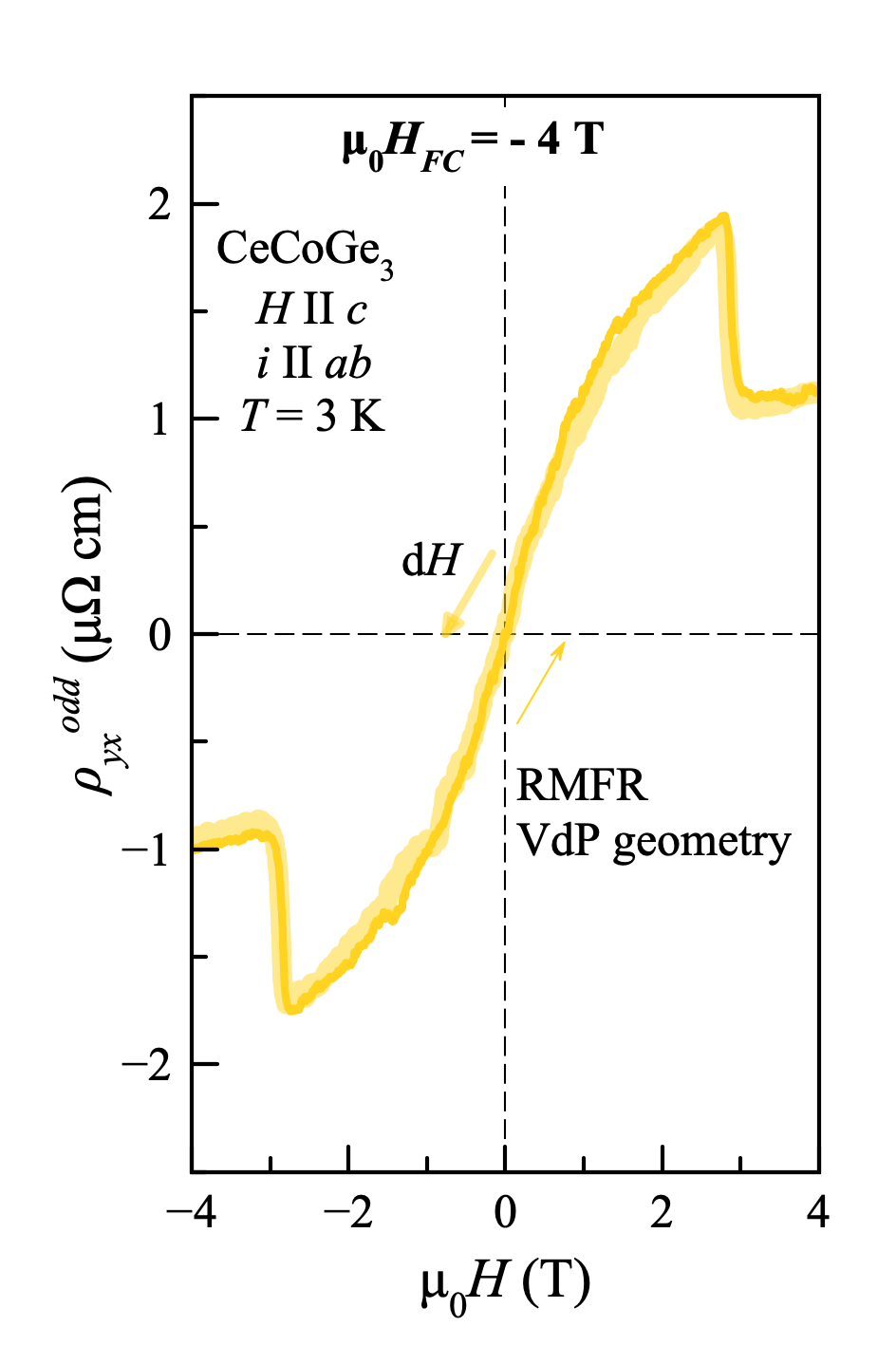}
\caption{\label{SI_RMFR} Isothermal Hall resistivity $\rho_{yx}^{odd}$ of CeCoGe$_3$ extracted using RMFR in a VdP geometry with Eq. ~\ref{Eq:spinningcurrents} on an independent crystal from the main text  measured at temperature $T~=$ 3 K with magnetic field $H\parallel c$ and current $I \parallel ab$ after field cooling the sample in $\mu_0H_{FC}~=~- 4$ T. }
\end{figure}

\begin{figure}
\centering
\includegraphics[width = 0.5\textwidth]{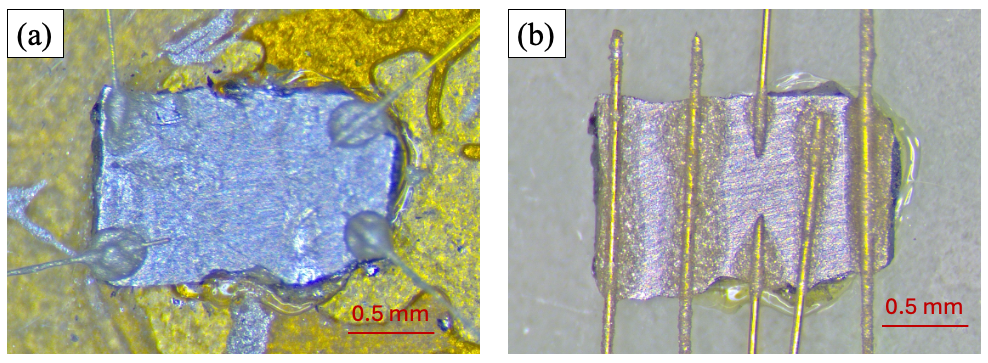}
\caption{\label{geometry} (\textbf{a}) Van der Pauw and (b) Hall bar geometry used for measuring the Hall effect of Co$_3$Sn$_2$S$_2$. }
\end{figure}

\end{document}